\documentclass[12pt, preprint]{aastex}
\usepackage{lscape}
\usepackage{subfigure}
\usepackage{graphicx}
\usepackage{multirow}
\usepackage{latexsym}
\usepackage{amssymb}
\usepackage{epsf}

\newcommand{\kms}{km s$^{-1}$ }
\newcommand{\kmsn}{km s$^{-1}$}

\newcommand{\his}{{\rm H\,}{{\sc i }}}

\newcommand{\CO}{$^{13}$CO }

\begin{document}
\title{The Turbulence Spectrum of Molecular Clouds in the Galactic Ring Survey: A Density-Dependent PCA Calibration }
\author{Julia~Roman-Duval\altaffilmark{1,2}, Christoph~Federrath\altaffilmark{3, 4}, Christopher~Brunt\altaffilmark{5}, Mark~Heyer\altaffilmark{6}, James~Jackson\altaffilmark{1}, Ralf S.~Klessen\altaffilmark{3}}
\altaffiltext{1}{Institute for Astrophysical Research at Boston University, 725 Commonwealth Avenue, Boston MA 02215; jduval@bu.edu, jackson@bu.edu}
\altaffiltext{2}{Space Telescope Science Institute, 3700 San Martin Drive, Baltimore, MD 21218; duval@stsci.edu}
\altaffiltext{3}{Zentrum f\"{u}r Astronomie der Universit\"{a}t Heidelberg, Institut f\"{u}r Theoretische Astrophysik, Albert-Ueberle-Str.~2, D-69120 Heidelberg, chfeder@uni-heidelberg.de, klessen@uni-heidelberg.de}
\altaffiltext{4}{Ecole Normale Sup\'{e}rieure de Lyon, CRAL, 69364 Lyon Cedex 07, Francef}
\altaffiltext{5}{School of Physics, University of Exeter, Stocker Road, Exeter EX4 4QL, UK}
\altaffiltext{6}{Department of Astronomy, University of Massachusetts, Amherst, MA 01003-9305, heyer@astro.umass.edu}
\begin{abstract}
\indent Turbulence plays a major role in the formation and evolution of molecular clouds. The problem is that turbulent velocities are convolved with the density of an observed region. To correct for this convolution, we investigate the relation between the turbulence spectrum of model clouds, and the statistics of their synthetic observations obtained from Principal Component Analysis (PCA). We apply PCA to spectral maps generated from simulated density and velocity fields, obtained from hydrodynamic simulations of supersonic turbulence, and from fractional Brownian motion fields with varying velocity, density spectra, and density dispersion. We examine the dependence of the slope of the PCA pseudo structure function, $\alpha_\mathrm{PCA}$, on intermittency, on the turbulence velocity ($\beta_v$) and density ($\beta_n$) spectral indexes, and on density dispersion. We find that PCA is insensitive to $\beta_n$ and to the log-density dispersion $\sigma_s$, provided  $\sigma_s$ $\leq$ 2. For $\sigma_s$ $>$ 2, $\alpha_{PCA}$ increases with $\sigma_s$ due to the intermittent sampling of the velocity field by the density field. The PCA calibration also depends on intermittency. We derive a PCA calibration based on fBms with $\sigma_s$ $\leq$ 2 and apply it to 367 \CO spectral maps of molecular clouds in the Galactic Ring Survey. The average slope of the PCA structure function, $\langle\alpha_\mathrm{PCA}\rangle=0.62\pm0.2$, is consistent with the hydrodynamic simulations and leads to a turbulence velocity exponent of $\langle\beta_v\rangle=2.06\pm0.6$ for a non-intermittent, low density dispersion flow. Accounting for intermittency and density dispersion, the coincidence between the PCA slope of the GRS clouds and the hydrodynamic simulations suggests $\beta_v$ $\simeq$ 1.9, consistent with both Burgers and compressible intermittent turbulence. 
\end{abstract}

\maketitle

\section{Introduction}

\indent Turbulence plays a major role throughout the entire lifetime of a molecular cloud, from its formation to its fragmentation and collapse into clumps, cores, and stars. Supersonic turbulence is not only responsible for most of the kinetic energy supporting clouds against gravity \citep{williams00},  but is also intrinsically linked to the formation and structure of molecular clouds \citep[e.g.,][]{maclow04, elmegreen04, audit05, heitsch05, vazquez06, vazquez07, dobbs06, tasker09, banerjee09, glover10, klessen10, federrath11}. Turbulent shocks create local, over-dense regions within molecular clouds that may collapse into sheets, filaments and pre-stellar cores under the effect of self-gravity, a phenomenon known as turbulent fragmentation. In this context, the length scale of the energy source that drives turbulence can explain differences in star-formation efficiency and type --- clustered versus isolated \citep{klessen00}. In addition, the Initial Mass Function (IMF), which describes the relative probability of stars with different masses when they form, may directly result from the magnitude and spatial scale of turbulent fluctuations \citep{padoan02, hennebelle08, hennebelle09}. Hence, most analytic models of star formation rely on the index of the turbulent energy spectrum (i.e., the Fourier spectrum of the velocity fluctuations), and on the Probability Distribution Function (PDF) of the turbulent density field \citep[e.g.,][]{elmegreen97, padoan02, krumholz05, hennebelle08, hennebelle09}.  The energy spectrum is defined as $E({\bf k})=4\pi k^2\left |{\bf \hat v}({\bf k})\right |^2$, where ${\bf \hat v}({\bf k})$ is the Fourier transform of the velocity field, and obeys a power law ($E(k)$ $\propto$ $k^{-\beta_v}$), where $\beta_v=5/3$ for incompressible (Kolmogorov) turbulence \citep{frisch95}, and $\beta_v=2$ for pressureless (i.e., highly supersonic) shock-dominated turbulence, also called Burgers turbulence \citep{burgers74, passot88}. Intermittency, characterized by fluctuations occurring sporadically (both spatially and temporally) in the turbulent flow, also affects the nature of the turbulent cascade from large to small scales. Numerical simulations show that intermittency is mainly caused by the interaction of strong shocks, causing rare, strong density enhancements. Manifestations of intermittency are observable in the PDF of the density field \citep{klessen00b, kritsuk2007,schmidt09,fed09b}, and in the index $\beta_v$ of the energy spectrum \citep{dubrulle94, she94, boldyrev02a, boldyrev02b, schmidt08}. \\
\indent While Kolmogorov (incompressible) turbulence does not seem appropriate for molecular clouds, in which turbulent motions are highly supersonic (hence compressible), the application of Burgers turbulence is limited to compressive fields for which $\nabla \times {\bf v} = 0$  (e.g., expanding shells). Three-dimensional numerical simulations of both decaying and driven turbulence, however, show that the ratio of compressive to solenoidal (for which $\nabla \cdot {\bf v} = 0$) velocity dispersion is $\gamma = \langle{\bf v_c^2}\rangle/\langle{\bf v_s^2}\rangle = 0.1$--1 \citep{porter98, porter99, fed09b}, where ${\bf v_c}$  and ${\bf v_s}$ are the compressive and solenoidal parts of the velocity field, respectively. An alternative, which includes intermittency and compressibility effects, was proposed on the basis of the \citet{she94} model by \citet{dubrulle94, boldyrev02a, boldyrev02b, schmidt08}. In this theoretical context, the turbulent energy cascade in the inertial range exhibits properties of Kolmogorov turbulence, while close to the dissipative range, intermittent shock structures resulting from compressibility effects start to dominate energy transfer mechanisms. Depending on the dimension of the dominant dissipative, intermittent structures, log-Poisson models predict $\beta_v = 1.74$--1.83.\\
\indent The type of turbulence (Kolmogorov, Burgers, log-Poisson), along with the density spectrum and density PDF, influence the resulting stellar mass spectrum,  the IMF \citep[e.g.,][]{padoan02, hennebelle08, hennebelle09}. While the density PDF is better constrained with extinction data \citep{cambresy99, ossenkopf01, brunt10b}, the density spectrum has been investigated both with FIR dust emission observations \citep[e.g.,][]{block10} and with \his 21 cm spectral line observations \citep{stanimirovic99, elmegreen01}. The index of the energy spectrum, $\beta_v$, is the subject of many spectral line studies of molecular clouds \citep{heyer97, brunt02b, ossenkopf02, brunt03, heyer04, hilyblant08}, including this work.  \citet{heyer97} proposed to use Principal Component Analysis (PCA) to determine the index of the energy spectrum of molecular clouds using molecular line observations. PCA provides pairs of spatial and velocity scales, or ``PCA pseudo structure functions'', which describe the amount of velocity fluctuations contained within an eddy of a given spatial scale. PCA pseudo structure functions thus provide a description of the turbulent energy cascade. Since knowing the turbulence energy spectrum of molecular clouds has great implications for our understanding of fragmentation and star formation, our ultimate goal is to derive the exponent of the velocity power spectrum from the analysis (e.g., via PCA) of spectral line maps. PCA pseudo structure functions, however, are based on emission, and hence on a complex convolution between the density, velocity, and temperature fields, while a complete description of a turbulent gas flow requires separate statistical descriptions of the velocity and density fields. For instance, turbulent velocity fluctuations can make two physically separated emitting elements at different distances from the observer seem to overlap in velocity space \citep{ballesteros02}. This would produce an element of double the emissivity, which could be interpreted as a single emitting element of double density or column density. Temperature fluctuations also produce emissivity variations that can be interpreted as column density variations. This is not only a limitation of PCA, but also of all observational studies aimed at deriving the energy spectrum from various mathematical tools, such as Centroid Velocity Increments \citep[see, e.g., ][]{hilyblant08, brunt04}, Velocity Channel Analysis \citep{lazarian00, esquivel03}, and Velocity Coordinate Spectrum \citep{lazarian06}. Unfortunately, while observations provide column densities and their PDF, the temperature, velocity, and volumetric density fields, however, are not directly observable. In order to determine the relation between the PCA pseudo structure function and the Fourier spectrum of the velocity field, we therefore need to disentangle the density, velocity, and temperature contributions to the observed position-position-velocity data and to the shape of the PCA pseudo structure function. This can be done by comparing the PCA pseudo structure function of simulated spectral maps obtained from simulations of molecular clouds to the statistics of the input velocity, density, and temperature fields. Here, we use isothermal simulations, and hence neglect the molecular excitation problem and the effects of star formation on the gas temperature to concentrate on density and velocity fluctuations.  Thus, our goal is to establish a {\it calibration} relation between the slope of the PCA pseudo structure function and the slope of the turbulence velocity spectrum, and to investigate how this calibration relation varies with properties of the density field. \\
\indent \citet{brunt03etal} established a calibration relation based on MHD simulations. Their simulations included many of the important physical processes in the ISM, such as gravity, magnetic fields, star formation feedback, heating and cooling.  The focus of their study was to link the PCA derived 
relationship between velocity differences and spatial scale to a particular order of structure function. They were able to show that PCA pseudo structure functions correspond to a low order ($\simeq$ 1) structure function even in the regime of 
strong intermittency. The resolution of their simulation was however too small to allow the existence of an inertial range, where the slope of the power-law velocity spectrum can be measured. As a result, their simulation had to be modified a posteriori in Fourier space to create a power-law velocity spectrum. In addition to this limitation, the turbulence in the MHD simulations by \citet{brunt03etal} were mostly driven compressively by expanding shells, occurring in star-forming regions.\\
\indent  In this paper, we investigate the statistical relation between the PCA pseudo structure function and the intrinsic index of the velocity field for two types of forcing --- compressive and solenoidal --- based on hydrodynamic simulations with a distinct inertial range \citep{fed09b}, combined with Fractional Brownian Motion (fBm) simulations. We examine the sensitivity of this relation to the statistics of the density field by varying both the exponent of the power-law density spectrum and the density PDF of the fBm simulations. Section~\ref{pca_section} describes PCA and the details of the method. Section~\ref{simulations_section} describes the simulations. In Sections~\ref{pca_hydro_section} and \ref{pca_fbm_section}, we present the results of PCA applied to the hydrodynamic and fBm simulations, and provide a PCA calibration in Section~\ref{calibration_lognormal_section}. We apply PCA to spectral observations of molecular clouds taken as part of the Galactic Ring Survey \citep{GRS} in Section~\ref{application_section}. Section~\ref{conclusion} consist of a brief conclusion. \\


\section{Principal Component Analysis (PCA)}\label{pca_section}

\subsection{Method}
\indent PCA, first suggested as a tool to derive energy spectra from spectral observations by \citet{heyer97},  detects line profile differences due to the turbulent nature of the flow as a function of  spatial scale. The line profile differences are represented by the eigenspectra. The spatial scales over which those line profiles differ are detected in the integrated intensity images (the principal components) of the eigenspectra. PCA thus provides pairs of spatial and velocity scales detected in a PPV cube, similar to a structure function, $\delta v = f(\delta \ell)$. This so-called PCA pseudo structure function describes the amount of kinetic energy contained within a given spatial scale. The details of the method have been described in \citet{brunt02a} and are  summarized below.\\
\indent The observational data obtained from spectral line mapping of molecular clouds can be represented as PPV cubes, which consist of an ensemble of $N_{r}$ spectra $T({\bf r}, v)$ =  $T(x_i, y_i, v_j)$ = $T_{ij}$ ($i$ = 1..$N_{r}$, $j$= 1..$N_v$) of a molecular spectral line at different positions ${\bf r}$ on the sky. $N_v$ is the number of velocity channels. The identified line profile differences are the eigenvectors (or eigenspectra) of the covariance matrix ${\bf S}$ \citep{heyer97, brunt02a}, such that ${\bf S}{\bf u}^{(n)}$ = $\lambda^{(n)}{\bf u}^{(n)}$, where $n$ is an index that labels the eigenvectors in order of decreasing corresponding eigenvalue  ($n$ = 1..$N_v$). The projection of the eigenvectors ${\bf u}^{(n)}$ onto the PPV cube ordered by decreasing eigenvalue $\lambda^{(n)}$ yields the $N_v$ principal components ${\bf PC}^{(n)}$  of the PPV cube, such that ${\bf PC}^{(n)}$ = $^t{\bf T}{\bf u}^{(n)}$ (of size $N_{r}$). \\
\indent The spatial and velocity scales characteristic of the principal components and eigenvectors are computed from their Auto-Correlation Function (ACF). The spatial scale $\delta \ell$ at which the ACF of the $n^{th}$ principal component falls by one $e$-fold defines the $n^{th}$ characteristic spatial scale. In a similar way, the corresponding $n^{th}$ velocity scale $\delta v$ is determined from the $e$-fold of the ACF of the $n^{th}$ eigenvector. In the end, PCA provides $N_v$ pairs of increasingly smaller spatial ($\delta \ell$) and velocity ($\delta v$) scales, similar to a structure function. This relationship can be approximated by a power law, $\delta v = v_0 \delta \ell^{\alpha_\mathrm{PCA}}$, where $v_0$ and $\alpha_\mathrm{PCA}$ are the amplitude and the slope of the PCA pseudo structure function. 

\subsection{Definition of spatial scales}
\indent The spatial and velocity scales are chosen to be the scales at which the normalized ACF falls by one $e$-fold. In numerous cases, the ACF of the 0$^{th}$ principal component does not fall below one $e$-fold due to the overall correlation of the bulk of the emission. In such instances, the 0$^{th}$ principal component was excluded from the PCA structure function. \\
\indent Furthermore, the determination of the velocity scales is straightforward because the ACF of the eigenvectors is one-dimensional. However, the ACF of the principal component images is two-dimensional. Because of the intrinsic shape of molecular clouds, the ACFs of the 1$^{st}$ and 2$^{nd}$ principal components are often elliptical, such that the spatial scales depend on the direction along which they are calculated. In order to measure spatial scales consistently, independent of the geometry of the cloud, the ACFs of the principal components were fitted to an ellipse. The ACFs were then rotated such that their long axes are horizontal, and the one $e$-fold spatial scales $\delta x$ and $\delta y$ were computed along the $x$ and $y$ cardinal directions. The spatial scales of the PCA pseudo structure function are then defined as $\delta \ell = \sqrt{(\delta x^2 + \delta y^2)/2}$, to stay consistent with previous PCA studies \citep[e.g.,][]{heyer97, brunt02a, brunt02b, brunt03etal}. 

\section{Simulations}\label{simulations_section}

\subsection{Hydrodynamic simulations}\label{hydro_sim_section}

\indent We numerically modeled isothermal driven turbulence on a periodic uniform grid with $1024^3$ grid cells. We refer the reader to \citet{fed08, fed09a, fed09b} for the details of the simulations. Two kinds of forcing were implemented: solenoidal (or divergence-free) forcing for which $\nabla\cdot{\bf f} = 0$, and compressive (curl-free) forcing for which $\nabla \times {\bf f} = 0$. These two types of forcing mimic actual mechanisms responsible for driving turbulence in the ISM. For instance, galactic shear corresponds to solenoidal forcing, while supernova explosions are a compressive way to drive turbulence. For each forcing case, 81 snapshots of the velocity and density fields, spanning eight large-scale turbulent crossing times were recorded in the regime of fully developed, supersonic turbulence. To facilitate the computation of structure functions and PCA, all snapshots were resampled on a $256^3$ grid. The resampling from $1024^3$ to $256^3$ did not affect the inertial range scaling \citep{fed09a}. The simulations, originally with a mean density of unity and an RMS Mach number, $\mathcal{M}=5.5$, were rescaled to a velocity standard deviation of 1$\,$\kms ($T=10\,$K, assuming that the gas is composed of pure molecular hydrogen) and a mean density of 500$\,$cm$^{-3}$.\\
\indent The statistics of the solenoidally and compressively forced simulations (in particular velocity and density power spectra and density PDFs) were computed in \citet{fed08, fed09a, fed09b} and are summarized in Table~\ref{stat_hydro_table}. The inertial range of the hydrodynamic simulations only extends between $k=5$ and $k=15$ due to numerical viscosity. In addition to the average statistics summarized in Table~\ref{stat_hydro_table}, we computed the inertial range exponent $\beta_v$ of the energy spectrum for each snapshot.  An important point for the upcoming analysis is that the density PDFs of the simulated fields are only approximately lognormal \citep{fed08,fed09b} due to significant intermittency. Deviations from a lognormal density PDF can be estimated via the skewness $\mathcal{S}_s$ and kurtosis $\mathcal{K}_s$ of the logarithm of the density, $s=ln(n/\langle n \rangle)$, where $\langle n \rangle$ in the mean density. For a perfectly lognormal distribution, $\mathcal{S}_s$ $=$ 0 and $\mathcal{K}_s$ $=$ 3. For the HD simulations, the deviations of the skewness and kurtosis from these fiducial values are higher in the compressive forcing case than in the solenoidal forcing case \citep[][and Table~\ref{stat_hydro_table}]{fed09b}. We characterize the width of the density PDF by the density dispersion, $\sigma_n/\langle n \rangle$, and the log-density dispersion, $\sigma_s$, which is the standard deviation of the logarithm of the density. The density PDF of the compressively forced field has a roughly three times higher standard deviation than the solenoidally forced counterpart at the same RMS Mach number, emphasizing the importance of studying different turbulent injection mechanisms. The values of $\sigma_n/\langle n \rangle$ and $\sigma_s$ for the HD simulations are also given in Table ~\ref{stat_hydro_table}.

\subsection{Fractional Brownian Motion}\label{fbm_section}

\indent The hydrodynamic simulations have a unique velocity power spectrum, which cannot be varied. In other words, the compressively and solenoidally forced hydrodynamic simulations each provide one point in the calibration relation. In contrast, we aim to establish a relation between the PCA pseudo structure function and the slope of the energy spectrum over a range of exponents for the velocity spectrum, which reflects different types of turbulence (e.g., compressible, incompressible, intermittent, non-intermittent, etc...) in the ISM. This can be accomplished by varying the exponent of the energy spectrum, of the density spectrum, and the dispersion of the density PDF  of {\it Fractional Brownian Motion} structures \citep[fBms, see, e.g.,][]{stutzki98}, and by comparing the energy spectrum of the fBms to the PCA pseudo structure function obtained from the corresponding simulated spectral map. \\
\indent A comprehensive study of fBms by \citet{stutzki98} summarizes the current knowledge of fBms, and we do not repeat it here. The details of the method we use to generate such fields are presented in \citet{ossenkopf06}, and are summarized by the following. An fBm can be generated in Fourier space by creating an isotropic amplitude following a power law, $A({\bf k})=A_0\,k^{-\gamma}$. A phase $\phi({\bf k})$ is randomly generated, using a uniform distribution between $-\pi$ and $\pi$ to obtain the final Fourier transform of the desired field (velocity or density), $\hat f({\bf k})$ $=$ $e^{i \phi({\bf k})}$. To ensure that the final field, obtained by taking the inverse Fourier transform of $\hat f({\bf k})$, is real, the condition $\phi({\bf k})=-\phi(-{\bf k})$ is imposed. \\
\indent Twelve velocity fields and fifteen density fields were created on a $257^3$ grid with a power-law Fourier spectrum of exponents ranging between $\beta_v=1.2$ and $\beta_v=3.4$ for the velocity field, and $\beta_n=0.6$ to $3.4$ for the density field.  The exponents of the density and velocity fields were varied independently. This range of values for the velocity spectrum covers different types of turbulence, including Kolmogorov, Burgers, and log-Poisson turbulence models. It also covers the case of systematic motions, such as infall, for which $\beta_v>3$ \citep{brunt02a}. Note that an amplitude $A_0=1$ was used, and the final velocity and density fields were rescaled to a velocity standard deviation of $1\,$\kms and a mean density of $500\,$cm$^{-3}$ a posteriori, as for the hydrodynamic simulations. For the density field, the rescaling is not as straightforward as for the velocity field, owing to its positivity. Hence, we subtracted from the original fBm its minimum value. The density field was then obtained by dividing the the fbm by its mean value and multiplying it by the desired mean density of 500 cm$^{-3}$. The density fBm fields created with this method approximately follow a Gaussian (also known as normal) distribution, with a standard deviation of $90$--$180\,\mathrm{cm}^{-3}$. The density dispersions of each fBm density field created with a gaussian PDF and a varying density spectrum are listed in Table \ref{table_density_disp}.  \\
\indent Simulations and observations show that the density PDF of isothermal supersonic turbulent flows is better approximated by a lognormal distribution \citep[i.e., a Gaussian distribution in the logarithm of the density, see][]{vazquez94, padoan97, passot98, fed08, price11}. Furthermore, we wish to examine the dependence of the calibration relation not only on the exponent of the density Fourier spectrum, but also on the density PDF. In order to create density fBms with a lognormal PDF of variable standard deviation, the method presented in \citet{ossenkopf06} and \citet{brunt02a} was followed. First, an fBm field was created with a power spectrum of slope $\beta_n=1$. This field represents the logarithm $\ln(n)$ of the desired density field $n$. The fBm field $\ln(n)$ was then rescaled given the desired mean $\langle n\rangle$ and standard deviation $\sigma_n$ of the desired density field. This rescaling is based on the relation between the mean and standard deviation of a lognormal field and its logarithm:

\begin{equation}\label{eq_ln_n1}
\langle\ln(n)\rangle = \ln(\langle n\rangle) - \frac{1}{2} \sigma_{\ln(n)}^2\,,
\end{equation}
\begin{equation}\label{eq_ln_n2}
\sigma_{\ln(n)} = \sqrt{\ln\left(1+\left(\frac{\sigma_n}{\langle n\rangle}\right)^2\right)}\;,
\end{equation}

\noindent where $\langle\ln(n)\rangle$ and $\sigma_{\ln(n)}$ are the mean and standard deviation of the logarithm of the desired field. Thus, if $\langle F \rangle$ and $\sigma_{F}$ are the mean and standard deviation of the fBm field $F = \ln(n)$, then $F$ is rescaled and exponentiated to produce the final density field with the desired lognormal distribution: 

\begin{equation}\label{rescaling_equation}
n = exp \left ((F- \langle F \rangle) \times \frac{\sigma_{\ln(n)}}{\sigma_F}  + \langle\ln(n)\rangle \right )
\end{equation}

\begin{figure}[ht]
\centering
\includegraphics[width = 9cm]{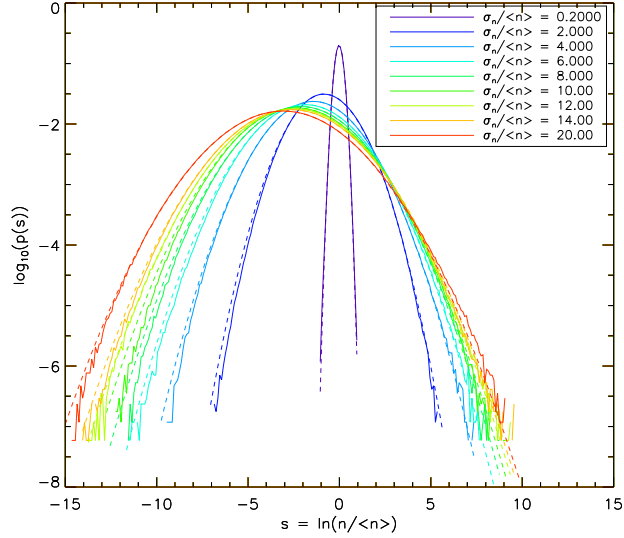}
\caption{PDFs, assumed to be lognormal, of the fBm density fields generated from exponentiation. The dashed lines represent the best lognormal fit to each PDF, the density dispersion of which is shown in the legend.  }
\label{fbm_pdfs}
\end{figure}

\noindent Six density fields with lognormal distributions were created, with a mean of $500\,\mathrm{cm}^{-3}$ and standard deviations of 100, 1000, 2000, 3000,  4000, 5000, 6000, 7000, and $10000\,\mathrm{cm}^{-3}$ ($\sigma_n/<\langle n \rangle$ $\simeq$ 0.2, 2, 4, 6, 8, 10, 12, 14, 20). Due to finite numerical resolution and low number statistics, the process of exponentiation may introduce small deviations from a lognormal PDF. To check the magnitude of the deviations of our fBms' density PDFs from a purely lognormal PDF, we computed the skewness and kurtosis of the logarithm of the density, which are listed in Table~\ref{table_beta_n}. The skewness is of the order of 0.01--0.04, so deviations from a purely lognormal distribution ($\mathcal{S}_s$ $=$ 0) are much smaller than for the HD simulations ($\mathcal{S}_s$ $=$ $-$0.1 and $-$0.26 from solenoidal and compressive forcing respectively). The kurtosis is also close (within 1.5\%) to the value of $\mathcal{K}_s$ $=$ 3 obtained from purely lognormal distributions. Last, exponentiation changes the index of the Fourier spectrum \citep{ossenkopf06}. The spectral indices of the density spectrum of the fBm created with a lognormal density PDF are listed in Table \ref{table_beta_n}. The PDFs of the fBms generated via exponentiation are shown in Figure~\ref{fbm_pdfs}. The dashed-lines represent the best lognormal fits to each PDF, the density dispersion of which is shown in the legend. 

\subsection{Generation of the spectral maps}\label{ppv_section}

\indent A simulated spectral map of the \CO emission line (i.e., a PPV cube) was created for each simulation assuming that the \CO line is optically thin, and assuming an abundance ratio $n(^{13}\mathrm{CO})/n(\mathrm{H}_2)=1.7\times10^{-6}$ \citep{LP90, BL87}. Note that  the abundance ratio  used in these simulations does not affect our results since it scales the CO intensity up or down but does not change the power spectrum of the density or velocity fluctuations. Thus, the simulated \CO PPV cubes were constructed from the density field $n(x, y, z)$ and velocity fields $v_x(x, y, z)$, $v_y(x, y, z)$, $v_z(x, y, z)$ using the following expression (which is an example along the $z$-direction) along each cardinal direction ($x$, $y$, $z$):

\begin{equation}
I_{\nu} ({\bf r}) =  \sum_z {j_{\nu}({\bf r}, u) dz}
\end{equation}

\noindent where {\bf r} = (x, y, z), $u$ is the velocity channel, $\alpha_\nu$ is the absorption coefficient at the center of the $^{13}$CO line, $B_\nu$ is the Planck function, and $I_\nu$ the specific intensity. The emissivity $j_\nu$  = $\alpha_\nu \: B_\nu$ has units of W/m/str/Hz and is expressed by:

\begin{eqnarray*}
j_{\nu}({\bf r}, u) dz & \; = \; & 3.6\times10^{-8}\exp\left(-\frac{5.28\,\mathrm{K}}{T({\bf r})}\right)
\nonumber \\
 & & \quad\times\;\frac{\mathrm{K}}{0.378\,T} \frac{n({\bf r})}{\mathrm{cm}^{-3}} \frac{dz}{\mathrm{pc}}\frac{c}{\sqrt{2\pi}\nu \sigma_v({\bf r})}
\nonumber \\
 & & \quad\times\;\exp\left[-(u - v_z({\bf r}))^2/2 \sigma_v({\bf r})^2\right]
\end{eqnarray*}

\noindent where $T$ is the temperature in K, $n$ the number density field in cm$^{-3}$, $v_z$ is the projection along the $z$-axis of the velocity vector ${\bf v}$ at position $(x,y,z)$ in $\mathrm{m}\,\mathrm{s}^{-1}$, and the velocity dispersion $\sigma_v$ in $\mathrm{m}\,\mathrm{s}^{-1}$, given by

\begin{equation}
\sigma_v({\bf r})^2 = \left( \frac{k\,T({\bf r})}{m_{\mathrm{CO}}} \right)^2 + \left(\frac{\partial v_z({\bf r})}{\partial z} dz\right)^2\;.
\end{equation}

\noindent In this expression, $m_{\mathrm{CO}}$ is the mass of the \CO molecule. The first term represents thermal motions, and the second gas flows. Note that both the fBms and the hydrodynamic simulations are isothermal, with a uniform temperature of $10\,\mathrm{K}$. In the following, we assume that the simulation box is $L$ $=$ 10 pc in size, and therefore, $dz$ $=$ $L/N$, where $N$ is the number of grid points on one side ($N$ $=$ 256 for the HD simulations, $N$ $=$ 257 for the fBm simulations). \\
\indent The previous equations are based on two limiting assumptions: 1) the CO line is optically thin, and 2) local thermodynamic equilibrium (LTE). While the latter is true for densities $>$ 100 cm$^{-3}$, CO is sub-thermally excited for smaller densities. Hence, the emission in the most diffuse regions of the simulations will be overestimated by our first assumption. On the other hand, CO becomes optically thick at column densities greater than $N(CO) > 10^{16}$ cm$^{-2}$, which is not accounted for by our simple radiative transfer model. Such column densities can be attained in the high density regions of the simulations for a reasonable cloud depth (a few pc). However the scaling of the column density is arbitrarily set by the choice of the size of the simulation box, so it is pointless to try to determine whether this limit is actually reached in the simulations. \\
\indent  The spectral maps resulting from hydrodynamic and fBm simulations were sampled on a 40 m s$^{-1}$ grid as in \citet{fed09b}. To test the effects of spectral resolution on the uncertainty in the exponent of the PCA pseudo structure functions derived for each snapshot of the HD simulations or fBm field, we produced spectral maps with spectral resolution 10 m s$^{-1}$ and 20 m s$^{-1}$. Increasing the spectral resolution to 20 m s$^{-1}$or 10 m s$^{-1}$ did not reduce the scatter in the exponent of the PCA pseudo structure function, and thus, we kept the original spectral resolution of 40 m s$^{-1}$ used in \citet{fed09b}.  A total of 486 \CO PPV cubes were generated from the hydrodynamic simulations (3 lines-of-sight directions for each of the 81 time snapshots for solenoidal and compressive forcing) and 255 \CO cubes were generated from the fBms (180 with variable density power spectra and 75 fBms with variable density PDFs). 

\section{PCA applied to hydrodynamic simulations}\label{pca_hydro_section}

\begin{figure}[ht]
\centering
\includegraphics[width = 9cm]{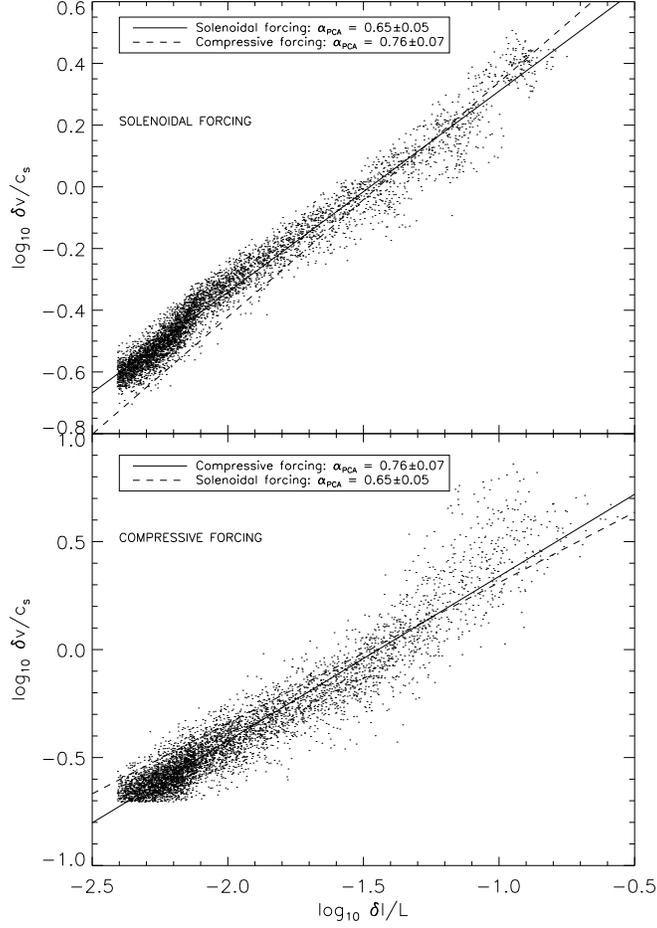}
\caption{PCA composite structure function for the hydrodynamic simulations, containing all spatial and velocity scales detected by PCA in all PPV cubes corresponding to solenoidal forcing (top) and compressive forcing (bottom). The solid line indicates the best fit. In the top panel (solenoidal forcing), the dashed line shows the best-fit from the compressive forcing case. In the bottom panel (compressive forcing), the dashed line shows the best-fit from the solenoidal forcing case.}
\label{plot_all_scales}
\end{figure}

\begin{figure}[p]
\centering
\subfigure{\includegraphics[width=8.5cm]{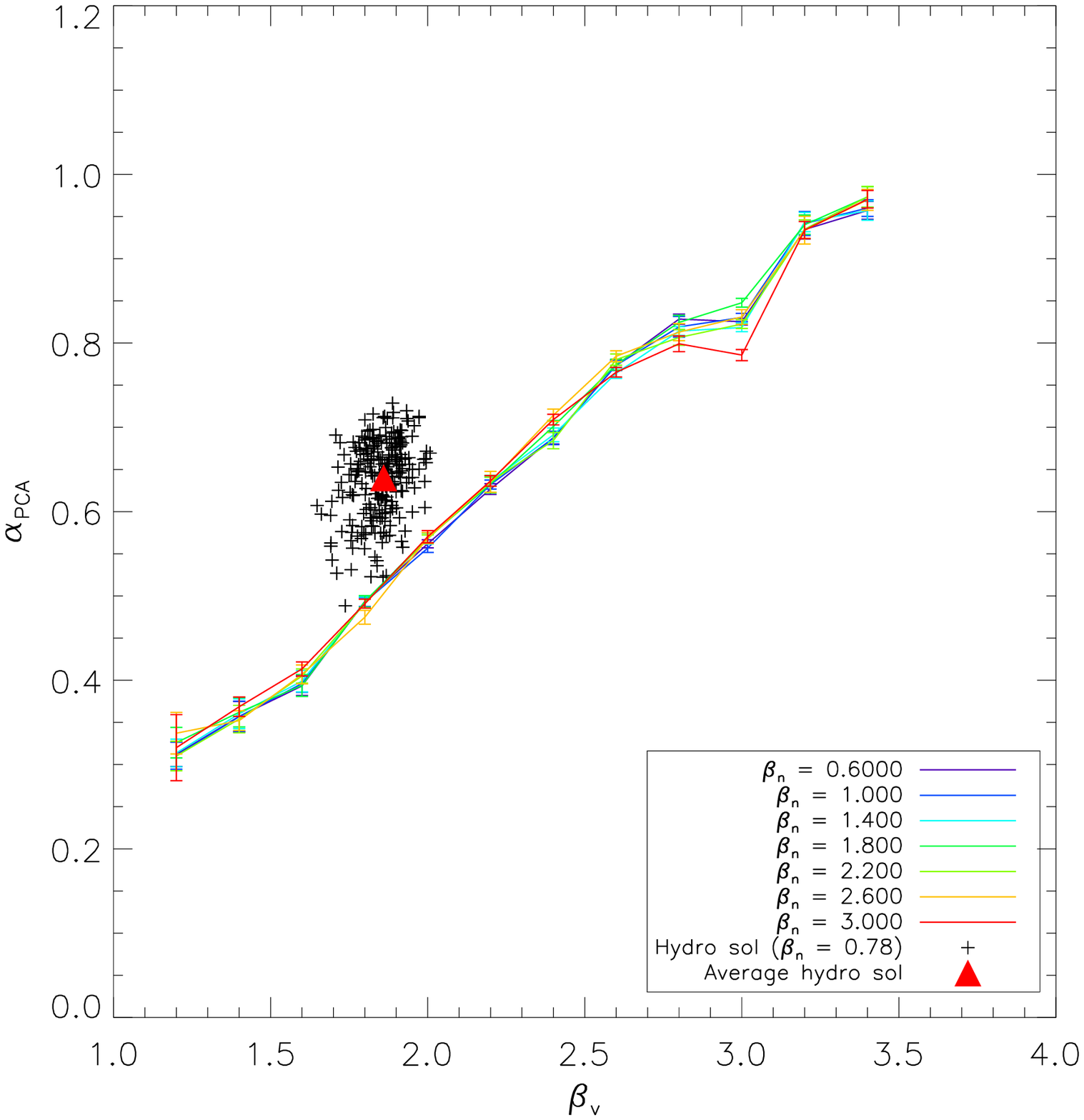}}
\subfigure{\includegraphics[width = 8.5cm]{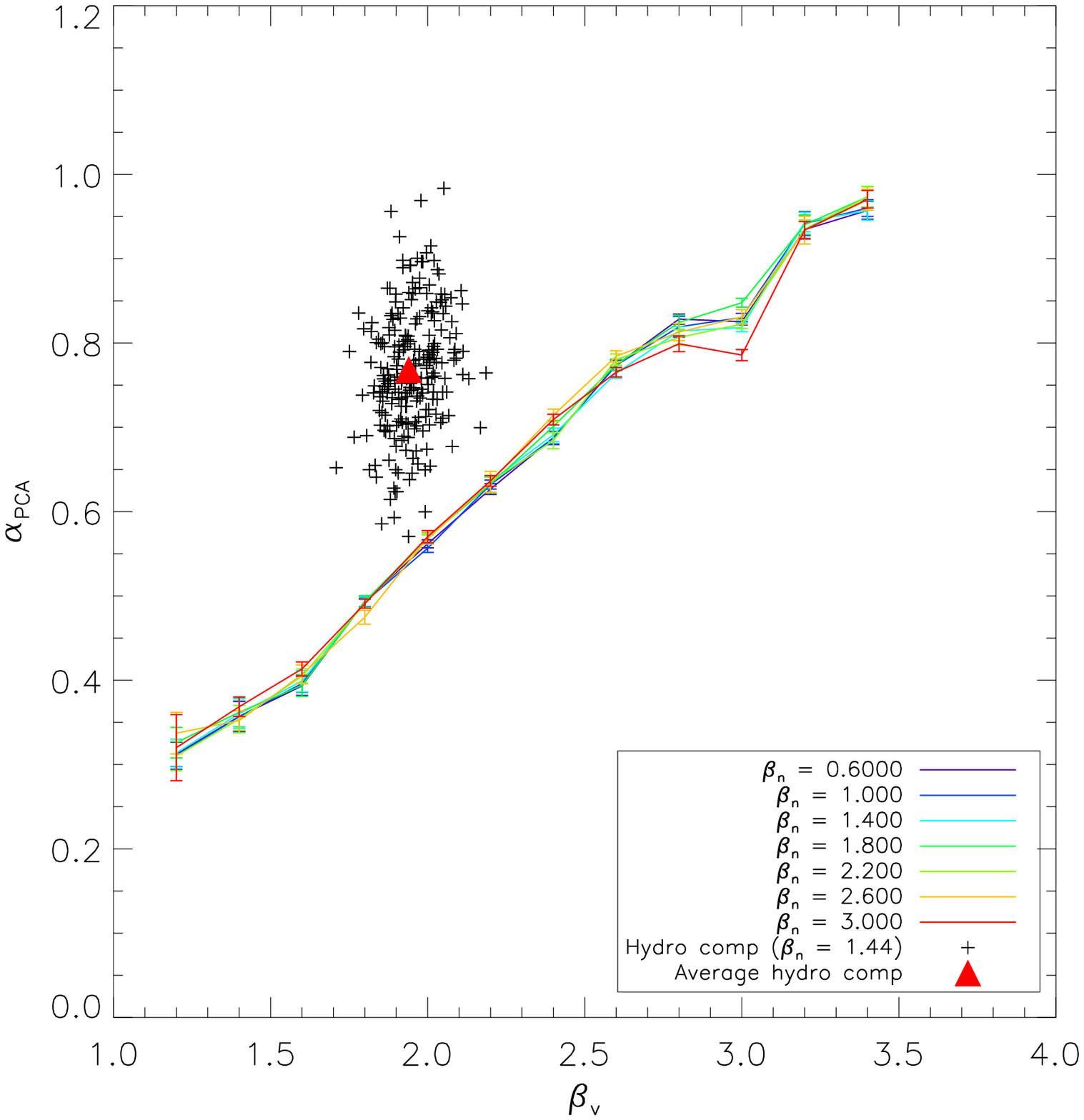}}
\caption{Calibration relation between the slope of the energy spectrum, $\beta_v$, and the slope of the PCA pseudo structure function, $\alpha_\mathrm{PCA}$, derived from PCA applied to fBms and hydrodynamic simulations. The top and bottom panels correspond to solenoidal and compressive forcing, respectively. The colored lines correspond to the relation between $\beta_v$ and $\alpha_\mathrm{PCA}$ obtained from fBms with different density spectra, indicated in the legend. The black crosses correspond to the calibration relation deduced from each hydrodynamic time snapshot, the average of which is shown by the red triangle. }
\label{calibration}
\end{figure}

\indent  PCA was applied to all 486 PPV cubes generated from hydrodynamic simulations and a power law was fitted to each resulting PCA pseudo structure function. \citet{fed09b} presented the time-averaged PCA pseudo-structure function for the solenoidally and compressively forced hydrodynamic simulations, with slopes $0.66\pm0.05$ and $0.76\pm0.09$ respectively. Here, we also derive the average slope of the PCA pseudo structure function, averaged over all time snapshots and all three lines of sight ($x$, $y$, and $z$)  and find $\langle\alpha_\mathrm{PCA}\rangle_\mathrm{sol}=0.64$ (standard deviation 0.05) and  $\langle\alpha_\mathrm{PCA}\rangle_\mathrm{comp}=0.77$ (standard deviation 0.07) for the solenoidally and compressively forced simulations, respectively. Typical errors on the slope of the PCA pseudo structure function for individual snapshots are 0.02 and 0.04 for solenoidal and compressive forcing respectively. In addition, Fig.~\ref{plot_all_scales} shows all the pairs of spatial and velocity scales detected by PCA in all the PPV cubes obtained from hydrodynamic simulations. A power-law fit to this composite PCA pseudo structure function yields an exponent of $\alpha_\mathrm{PCA}^\mathrm{sol}=0.65\pm0.05$ and $\alpha_\mathrm{PCA}^\mathrm{comp}=0.76\pm0.07$ for the solenoidally and compressively forced simulations, respectively. These results are in very good agreement with the slope of the time-averaged PCA pseudo structure function from \citet{fed09b}. The results of PCA applied to individual time snapshots of the hydrodynamic simulations are shown as black crosses in Figure~\ref{calibration}, their average being indicated by a red triangle. \\

\section{PCA applied to fBms: Sensitivity of the PCA calibration to the density spectrum}\label{pca_fbm_section}

\indent PCA was applied to the 180 PPV cubes generated from fBms velocity and density fields with varying power-law Fourier spectra. The colored lines in Figure~\ref{calibration} show the slope of the PCA pseudo structure function as a function of the exponent of the velocity spectrum for different density power spectra (the exponent of which is indicated in the legend).  \\
\indent The slope of the PCA pseudo structure function, $\alpha_\mathrm{PCA}$, increases with $\beta_v$, in agreement with previous calibrations based on fBms shown by the dashed line \citep{brunt02a}. The variation of $\alpha_\mathrm{PCA}$ with $\beta_v$ is independent of the exponent of the density Fourier spectrum over the range $\beta_n=0.6$ to $\beta_n=3.4$. In addition, the average relation between $\alpha_\mathrm{PCA}$ and $\beta_v$ obtained from HD simulations is too high (by a factor 2-3$\sigma$) compared to the relation obtained from fBms with gaussian PDFs. \\
\indent  There are two major differences between the HD simulations and the fBms: 1) the HD simulations are intermittent, while the fBms are not; and 2) the fBms and HD density fields have different density PDFs, both in shape and standard deviation. The fBms have a gaussian density PDF of density dispersion $\sigma_n/\langle n \rangle$ $\simeq$ 0.2-0.3, while \citet{fed08, fed09b} showed that the density PDFs of the HD simulations approximately follow lognormal distributions (i.e., gaussian in the logarithm of n).  The discrepancy between the PCA pseudo structure functions obtained from HD simulations and fBms therefore suggests that the PCA calibration depends on the level of intermittency (both in the velocity and density fields), on the shape of the density PDF, and its density dispersion. This hypothesis is further explored in the next section.

\section{Sensitivity of the PCA calibration to intermittency and to the density PDF}\label{calibration_lognormal_section}

\subsection{PCA applied to fBms with lognormal density PDFs of varying dispersion}
\indent We have further tested the variation of the PCA calibration relation with density PDF by applying PCA to 180 PPV cubes generated from fBms with lognormal density PDFs of varying standard deviations, ranging from $100$ to $10000\,\mathrm{cm}^{-3}$ ($\sigma_n/\langle n \rangle$ $=$ 0.2--20, $\sigma_s$ $=$ 0.2--2.45). Since the fBm velocity fields are not intermittent, we can thus isolate the effects of the density dispersion independent of the effects of intermittency in the velocity field, manifest in the HD simulations. The fBm density fields with varying density PDFs were created according to the method described in Section~\ref{fbm_section}. Although a constant spectral index $\beta_n$ $=$ 1 characterizes the logarithm of the density fields, exponentiation changes the power spectrum \citep{ossenkopf06}. As a result, the spectral indices of the fBm density fields with lognormal PDFs are not equal to $\beta_n$ $=$ 1, and are listed in Table \ref{table_beta_n}. Nonetheless, we have shown that the calibration relation is insensitive to the index of the density spectrum in the previous section. Thus, the variations of $\beta_n$ in the density fields with lognormal PDFs should not  cause any variations in the calibration relation. \\
 \indent The colored lines in Figure~\ref{calibration_lognormal} show the PCA calibration obtained for lognormal density PDFs of varying standard deviation (indicated in the legend). At high $\beta_v$, the PCA calibration becomes unstable because it depends very strongly on a few Fourier components. We do not take into account values of $\beta_v$ $>$ 2.6 in the following. For $\sigma_s$ $\leq$ 2, we do not find any significant variation in the PCA calibration as a function of log-density dispersion, while there is a sudden increase in $\alpha_{PCA}$ and its scatter for a given $\beta_v$ for $\sigma_s$ $>$ 2.  Actually, the PCA calibration becomes quite unstable for $\sigma_s$ $>$ 2. For $\sigma_s$ $\leq$ 2, the average calibration is shown by the black solid line. We derive a linear fit to the average PCA calibration obtained from fBms with lognormal density PDF of dispersion $\sigma_s$ $\leq$ 2, valid in the range $\beta_v$ $=$ 1.2-2.6: 
 
 \begin{equation}\label{calibration_equation}
 \beta_v = 0.20\pm0.05 + (2.99\pm0.09)\alpha_{PCA}
 \end{equation}
 
\noindent This calibration is essentially identical to the relation derived in \citet{brunt02a}, within the errors. \\
\indent The increase and instability in $\alpha_{PCA}$ for $\sigma_s$ $>$ 2 is likely due to the inability of the density field to properly sample the velocity field at such high density dispersion. Extreme density fluctuations intermittently sample the velocity field, producing an effect similar to intermittency in the velocity field itself, similarly to discontinuous velocity jumps. In fact, we attempted to perform a PCA run on an fBm with $\sigma_s$ $=$ 3, but the field was so extreme that no scales could be detected in the PCA pseudo structure function. Note that, although the density dispersion of the compressively forced simulations ($\sigma_s^{\mathrm{comp}}$ $=$ 3.04) is higher than the density dispersion of the fBm with $\sigma_s$ $=$ 2.45, this effect is not as strong for the HD fields because a high dispersion exponentiated fBm field is dominated by a few very high density point-like structures, while the hydrodynamically-produced density fields are dominated by a collection of filament-like structures. The latter are more spatially coherent, and capable of ($\simeq$uniformly) sampling the velocity field across a longer region of space than the former. This effect, examined in more details in the next section, also appears to be a threshold effect, as shown by the absence of significant variations in the PCA calibration for $\sigma_s$ $\leq$ 2.  
 
 \subsection{Effects of the density dispersion}\label{density_pdf_effects}
\begin{figure}
\centering
\includegraphics[height = 9cm]{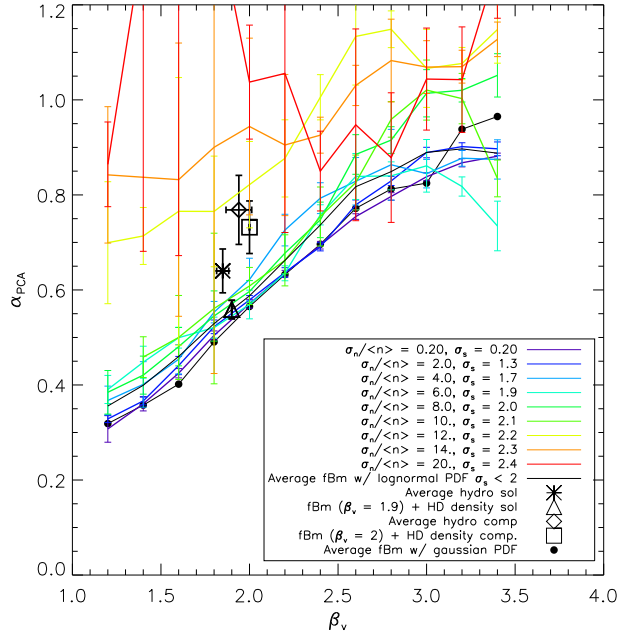}
\caption{Calibration relation obtained from hydrodynamical simulations and density fBm fields with lognormal distributions. The black star and triangle represent the average result of PCA applied to each time snapshot of the solenoidally and compressively forced  simulations respectively. The error bar represents the 1-$\sigma$ dispersion. The colored lines represent the calibration obtained from each fBm, with density PDF of standard deviation indicated in the legend.  }
\label{calibration_lognormal}
\end{figure}

\begin{figure}
\centering
\includegraphics[height = 9cm]{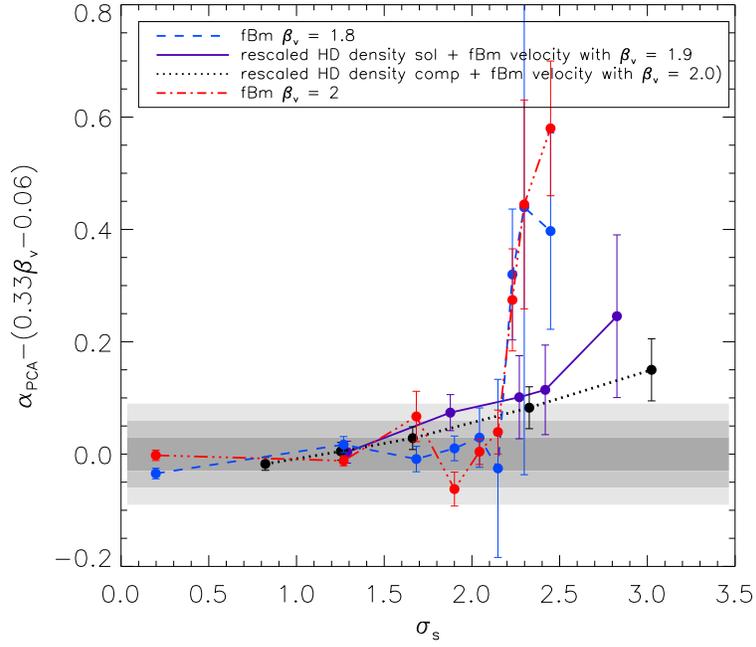}
\caption{Difference between the PCA slope $\alpha_{PCA}$ and the PCA slope $\alpha_{PCA}^{\mathrm{cal}}$ predicted by the calibration derived from fBms with $\sigma_s$ $\leq$ 2 as a function of log-density dispersion $\sigma_s$. The trends were obtained from the fBms with lognormal density PDFs and from spectral maps generated from rescaled HD density fields combined with fBm velocity fields (see Section \ref{density_pdf_effects}). The shaded area represents the 1, 2, 3$\sigma$ uncertainty in $\alpha_{PCA}^{\mathrm{cal}}$ from darkest to lightest. }
\label{plot_alpha_sigma_n}
\end{figure}

\indent The PCA calibration derived from fBms with lognormal density PDFs of varying density dispersion exhibits some dependency to the log-density dispersion, $\sigma_s$, but only above the threshold $\sigma_s$ $>$ 2 (see Fig. \ref{calibration_lognormal}).  The comparison between the PCA calibration derived from fBms with $\sigma_s$ $\leq$ 2 and HD compressively forced simulations, which have a log-density dispersion $\sigma_s^{\mathrm{comp}}$ $=$ 3.04 above the $\sigma_s$ $\simeq$ 2 threshold, also supports the hypothesis that the PCA slope depends on $\sigma_s$ for a given $\beta_v$. Indeed, in Fig. \ref{calibration_lognormal}, the PCA slope of the compressively forced simulations, shown as an open diamond, is $\alpha_{PCA}$ $=$ 0.76$\pm$0.07, while the calibration relation obtained from fBms of density dispersion below the threshold $\sigma_s$ $\leq$ 2 predicts $\alpha_{PCA}$ $=$ 0.58$\pm$0.03 for the corresponding $\beta_v$ $=$ 1.94$\pm$0.05, or a factor of 3$\sigma$ difference. In order to prove that this difference in PCA slope is indeed due to the higher density dispersion of the HD compressively forced density field, we have performed two tests.\\
\indent  First, we have generated spectral maps with the non-intermittent fBm velocity field of velocity spectrum $\beta_v$ $=$ 2 and the 81 snapshots of the compressively forced HD density field.  In this manner, we can isolate the effects of the HD density field from the effects of intermittency in the HD velocity field.  The average PCA slope of all snapshots is shown as an open square in Fig. \ref{calibration_lognormal}. This PCA run with the fBm velocity field and the HD density field can reproduce, well within the errors, the PCA slope of the spectral maps generated from HD density and velocity fields with compressive forcing (open diamond), and demonstrates that the log-density dispersion of the HD density field contributes significantly to the increase in PCA slope compared to the average PCA calibration obtained from fBms with $\sigma_s$ $\leq 2$. \\
\indent Second, we have rescaled both the HD compressively and solenoidally forced density fields to several log-density dispersions. For the compressively forced simulations, which originally have a density dispersion above the variation threshold of $\sigma_s$ $\simeq$ 2, we have scaled the log-density dispersion down to values ($\sigma_n/\langle n \rangle$ $=$ 0.8, 1.4, 2.0, 3.4, or $\sigma_s$ $=$ 0.8, 1.3, 1.7, 2.3) below or around this threshold in order to test if the PCA slope can be decreased down to values consistent with the PCA calibration obtained from lognormal fBms with $\sigma_s$ $\leq$ 2. For the solenoidally forced simulations, which originally have a density dispersion below the variation threshold of $\sigma_s$ $\simeq$ 2, we have scaled the log-density dispersion up to values ($\sigma_n/\langle n \rangle$ $=$ 3.6, 5.9, 7.1, 11.8, or $\sigma_s$ $=$ 1.9, 2.3, 2.4, 2.8) around or above this threshold. In this case, we thus test whether, according to expectations, the PCA slope obtained from the HD rescaled density fields with solenoidal forcing increases up to the level of the original compressively forced HD simulations for $\sigma_s$ above the variation threshold of the log-density dispersion. \\
\indent The rescaling was done similarly to the fBms with lognormal density PDF described in Equation \ref{rescaling_equation}, i.e. by rescaling the log of the density ($F$ in Equation  \ref{rescaling_equation}) to the desired mean and dispersion calculated from Equations \ref{eq_ln_n1} and \ref{eq_ln_n2} and from the desired rescaled $\sigma_n/\langle n \rangle$. We then exponentiated the rescaled log-density field. Because the HD density fields deviate from a lognormal distribution, the resulting log-density dispersion $\sigma_s$ is not exactly related to the input $\sigma_n$ by Equations \ref{eq_ln_n1} and \ref{eq_ln_n2}, but these equations nonetheless provide a good approximation. The exact rescaled values of $\sigma_n$ and $\sigma_s$ cited above are derived directly from the rescaled HD density fields. \\
\indent Finally, we applied PCA to the spectral maps generated from the rescaled HD density fields and the non-intermittent fBm velocity field with $\beta_v$ $=$ 1.9 for the solenoidally forced density field, and $\beta_v$ $=$ 2 for the rescaled, compressively forced HD density field. Again, the HD rescaled density fields are combined with non-intermittent fBm velocity fields in order to isolate the effects of the density dispersion from the effects of intermittency in the HD velocity field, which can potentially affect the calibration, as shown in Section \ref{operating_order}. Fig. \ref{plot_alpha_sigma_n} shows the variations of the difference between the resulting PCA slope, $\alpha_{PCA}$, and the PCA slope predicted from the calibration obtained with fBms of density dispersion $\sigma_s$ $\leq$ 2, $\alpha_{PCA}^{\mathrm{cal}}$ (see Equation \ref{calibration_equation}), as a function of $\sigma_s$. In this Figure, we have also included the variations of $\alpha_{PCA}-\alpha_{PCA}^{\mathrm{cal}}$ obtained from fBms with $\beta_v$ $=$ 1.8 and $\beta_v$ $=$ 2.0. The shaded area indicates the 1, 2, 3 $\sigma$ uncertainty in $\alpha_{PCA}^{\mathrm{cal}}$ from darkest to lightest. Note that  $\alpha_{PCA}-\alpha_{PCA}^{\mathrm{cal}}$ is not identically zero for fBms with $\sigma_s$ $\leq$ 2 because the calibration derived in Equation \ref{calibration_equation} is a fit to the average trend obtained from fBms with $\sigma_s$ $\leq$ 2. Fig. \ref{plot_alpha_sigma_n} demonstrates that, for $\sigma_s$ $\leq$ 2, the calibration is constant with $\sigma_s$ within the errors, while $\alpha_{PCA}$ starts to increase and deviate significantly from the calibration for $\sigma_s$ $>$ 2 due to the poor sampling of the velocity field by the density field. Above $\sigma_s$ $\simeq$ 2, the variations of $\alpha_{PCA}$ with $\sigma_s$ for a given $\beta_v$ are steeper and more uncertain for the fBms than for the HD rescaled fields, which is also seen in Fig. \ref{calibration_lognormal}. As mentioned before, a probable explanation for this difference is that  the fBms are just dominated by a few point-like structures that sample the velocity field very poorly, while the HD density field structure consists of filament-like features which are more spatially coherent than the fBm density extrema. \\
 
 \subsection{Effects of intermittency in the velocity field and the operating order of PCA}\label{operating_order}
 \indent Although the log-density dispersion of the solenoidal forced simulations ($\sigma_s^{\mathrm{sol}}$ $=$ 1.3) is below the threshold of $\sigma_s$ $\simeq$ 2 above which $\alpha_{PCA}$ starts to increase with $\sigma_s$, Figure \ref{calibration_lognormal} shows that the PCA slope of the solenoidally forced HD simulations still stands out as being too high compared to the calibration derived from fBms. Indeed, the average slope of the PCA pseudo structure function is 0.65$\pm$0.05, while the average calibration obtained from fBms with density dispersion $\sigma_s$ $\leq$ 2 predicts  $\alpha_{PCA}$  $=$ 0.55$\pm$0.03 for the corresponding $\beta_v$ $=$ 1.86$\pm$0.05. In order to determine the cause of this discrepancy, we have first confirmed that the HD solenoidally forced density field was not causing this difference by applying PCA to spectral maps generated from a non-intermittent fBm velocity field of velocity spectrum $\beta_v$ $=$ 1.9 and the 81 snapshots of the solenoidally forced HD density field. The result is shown by the open triangle in Fig. \ref{calibration_lognormal} (this point is also shown in Fig. \ref{plot_alpha_sigma_n}). The average PCA slope of the spectral maps generated from the solenoidally forced HD density fields and fBm velocity field with $\beta_v$ $=$ 1.9 is in nearly perfect agreement with the PCA calibration derived from fBms velocity and density fields with lognormal density PDF of dispersion $\sigma_s$ $\leq$ 2. In Fig. \ref{plot_alpha_sigma_n}, the combination of the HD solenoidally forced density field and the fBm velocity field with $\beta_v$ $=$ 1.9 fits well within the errors in the trend $\alpha_{PCA}-\alpha_{PCA}^{\mathrm{cal}}$ versus $\sigma_s$. This demonstrates that the HD solenoidally forced density field is not causing the PCA slope obtained for HD solenoidally forced simulations to be too high compared to the calibration obtained for the same range of log-density dispersion. We conclude that, in this case, this discrepancy must be due to the intermittent structure of the HD velocity field.  \\ 
\indent \citet{brunt03etal} showed that the slope of the PCA pseudo structure function reflects the slope $\zeta_1$ of the first order structure function, defined as $\mathrm{SF}_1(\ell) = \langle v(r + \ell) - v(r) \rangle$, and that the calibration between $\zeta(1)$ and $\alpha_{PCA}$ is insensitive to the level of intermittency. On the other hand, they showed that the relation between $\alpha_{PCA}$ and the slope $\zeta(2)$ of the second order structure function, defined as $\mathrm{SF}_2(\ell) = \langle \left [ v(r + \ell) - v(r) \right ]^2 \rangle$, or equivalently the slope of the velocity spectrum $\beta_v$ $=$ 2$\zeta(2)+1$, depends on the level of intermittency. It was demonstrated theoretically by \citet{boldyrev02b} and numerically by \citet{schmidt08} that the relation between the order $p$ of the structure function $SF_p(\ell) =  \langle \left [ v(r + \ell) - v(r) \right ]^p \rangle$ and its exponent $\zeta(p)$ is more concave as the level of intermittency in the velocity field increases (i.e., it rises slower than linear). Thus, the ratio of $\zeta(1)$/$\zeta(2)$ increases as velocity fields becomes more intermittent. Because PCA traces $\zeta(1)$, it is therefore expected that  $\alpha_{PCA}$ increases with intermittency in the velocity field for a given $\beta_v$. This is confirmed by the fact that the intermittent HD solenoidally forced velocity field (rather than the density field) is causing the increase in $\alpha_{PCA}$ compared to the fBms of same $\beta_v$ (and log-density dispersion), which are not intermittent. In the following, we quantify this effect for the HD simulations.\\
\indent  The level of intermittency in the velocity field increases from fBms (non-intermittent), to solenoidally forced, to compressively forced HD simulations, as demonstrated in \citet{schmidt08}. As a result, the ratio $\zeta(1)/\zeta(2)$ is lowest for the fBms \citep[which are not intermittent and for which $\zeta(p)$ is linear with $p$, see][]{brunt03etal}, increases for solenoidally forced HD simulations, and increases even more for compressively forced simulations. We can estimate the ratio $\zeta(1)/\zeta(2)$ using the results from \citet{schmidt08}, who found $\zeta(1)/\zeta(3)$  $=$ 0.47$\pm$0.01 and $\zeta(2)/\zeta(3)$ $=$ 0.79$\pm$0.01 for the case of solenoidal forcing, and $\zeta(1)/\zeta(3)$  $=$ 0.63$\pm$0.01 and $\zeta(2)/\zeta(3)$ $=$ 0.90$\pm$0.01 for the compressively forced HD simulations. We can then derive $\zeta(1)$ for each forcing case using the following Equation:

\begin{equation}
\zeta(1) = \left [ \frac{\zeta(1)}{\zeta(3)} \right ] \left [\frac{\zeta(3)}{\zeta(2)} \right ] (\beta_v -1)
\end{equation}

\noindent where we used the fact that $\zeta(2)$ $=$ $\beta_v -1$. We find $\zeta(1)$ $=$ 0.51$\pm$0.03 for the solenoidal forcing case and $\zeta(1)$ = 0.66$\pm$0.04 for the compressive forcing case. Using Equation 31 in \citet{brunt03etal}, which relates  $\zeta(1)$ and $\alpha_{PCA}$, with $\gamma$ $=$ $\zeta(1)$ (the operating order of PCA), we predict $\alpha_{PCA}$ $=$ 0.62$\pm$0.03 for solenoidal forcing and $\alpha_{PCA}$ $=$ 0.75$\pm$0.07 for  compressive forcing. This prediction, based on the relation between $\zeta(1)$ and $\alpha_{PCA}$ established in \citet{brunt03etal} and the characterization of intermittency by the $\zeta(p)$ relation in the HD velocity field, is consistent within the error bars with the PCA slopes derived here for the HD simulations ($\alpha_{PCA}^{\mathrm{sol}}$ $=$ 0.65$\pm$0.05 and $\alpha_{PCA}^{\mathrm{comp}}$ $=$ 0.76$\pm$0.07).\\
\indent The difference between $\alpha_{PCA}$ obtained for HD simulations and fBms of similar log-density dispersion and velocity spectrum (see Fig. \ref{calibration_lognormal}) therefore confirms that intermittency in the velocity field changes the relation between $\beta_v$ and $\alpha_{PCA}$. For a given $\beta_v$, the variations of $\alpha_{PCA}$ between HD and fBms simulations can then be simply explained in terms of 1) the operating order of PCA and 2) the variations of the relation $\zeta(p)$ between the exponents of the structure functions of different orders $p$ with intermittency in the velocity field. In fact, one goal of this paper is precisely to investigate how the calibration between $\alpha_{PCA}$ and $\beta_v$ varies with intermittency and how to account for it in the derivation of $\beta_v$ from PCA. For the solenoidal forcing case, the difference in PCA slope between non-intermittent fBm and intermittent HD fields is entirely due to intermittency in the HD velocity field, since, in this case, the log-density dispersion of the density field is below the $\sigma_s$ $\simeq$ 2 threshold for which poor sampling of the velocity field causes the PCA slope to increase at constant $\beta_v$. For the compressive forcing case, we have demonstrated that the lack of sampling of the velocity field by the density field due to the high log-density dispersion contributes significantly to the difference in PCA slope between the compressively forced HD simulations and the fBms of same velocity spectrum. Nonetheless, the contribution from intermittency in the velocity field likely plays an important role, although it is difficult to detect it considering the larger error bars obtained for the HD simulations with compressive forcing. \\

\subsection{Consequence for molecular cloud observations}
\indent Applying a PCA calibration obtained from the idealized structures that are fBms to actual molecular clouds with high levels of intermittency and very high density dispersions will lead to an overestimation of $\beta_v$. For instance, in the case of the solenoidally forced simulations ($\alpha_{PCA}$ $=$ 0.65$\pm$0.05, $\mathcal{S}_s$ $=$ $-$0.10$\pm$0.11, $\mathcal{K}_s$ $=$ 3.03$\pm$0.17), the PCA calibration obtained from fBms with $\sigma_s$ $\leq$ 2 predicts $\beta_v$ $=$ 2.12$\pm$0.17, while the actual value of $\beta_v$ measured in the simulated velocity field is $\beta_v$ $=$ 1.86$\pm$0.05. For the compressively forced simulations ($\alpha_{PCA}$ $=$ 0.76$\pm$0.07, $\mathcal{S}_s$ $=$ $-$0.26$\pm$0.20, $\mathcal{K}_s$ $=$ 2.91$\pm$0.43), the PCA calibration obtained from fBms estimates $\beta_v$ $=$ 2.48$\pm$0.27, while actually, $\beta_v$ $=$ 1.94$\pm$0.05. The difference and induced uncertainty in these values of $\beta_v$ appears very large compared to the subtle difference between, e.g., Kolmogorov turbulence ($\beta_v$ $=$ 5/3) and Burgers turbulence ($\beta_v$ $=$ 2). As a result, one should explore values of $\beta_v$ implied by molecular line observations based on  PCA calibration relations derived for different density dispersions and levels of intermittency in the velocity field.

\section{Application to molecular clouds identified in the Galactic Ring Survey}\label{application_section}
In this section, we apply PCA to spectral maps of 750 molecular clouds identified in the Five College Radio Astronomy Observatory (FCRAO) Galactic Ring Survey \citep[GRS, see][]{GRS} for which distances are available from \citet{RD2009}. Several observational difficulties must first be overcome before PCA can be applied to observations of molecular clouds.

\subsection{Subtraction of the noise ACF}
\indent The noise inherent to observations of molecular clouds contributes to the observed ACF of the principal components. It can be shown  that the noise in the principal components is identical to the noise in the data \citep{brunt02a}. For spectroscopic data cubes obtained from on-the-fly mapping (as in the GRS), the noise is correlated between positions observed with the same reference position. As a result, the ACF of the noise contributes a powerful peak at the zero-lag, and other peaks corresponding to different correlation lengths of the noise \citep[see also][]{GRS}. The ACF of the noise therefore needs to be subtracted from the observed ACF of the principal components. The contribution of the noise to the ACF of the principal components was estimated by averaging the ACF of the 5 highest-order principal components, which should only contain contributions from noise, because they correspond to features of spatial scales well below the resolution limit. The ACF of the noise was then subtracted from the ACFs of all the principal components before computing the spatial scales.

\subsection{Effects of the finite telescope resolution on the determination of spatial scales}\label{beam}


\indent The convolution of the observed \CO data with the beam of the telescope affects the determination of spatial scales. The removal of the beam contribution to the observed spatial scales has been established by \citet{brunt99, brunt02a}, using approximations. Here, we analytically derive the exact scale correction for beam smearing, and test it using the HD numerical simulations. Let ${\widetilde X}$ be a two-dimensional image observed through a telescope with a two-dimensional beam B, and let $X$ be the ``true'' image (i.e., not convolved with the telescope beam). ${\widetilde X}$ is the convolution of the true image with the telescope beam: ${\widetilde X} = X \star B$. It can be demonstrated \citep[see, e.g., ][]{brunt99} that the ACF of the beam-smeared principal components, $\mathrm{ACF}_{\widetilde{\bf PC}^{(n)}}$,  is related to the ACF of the true principal components, $\mathrm{ACF}_{{\bf PC}^{(n)}}$, by $\mathrm{ACF}_{\widetilde{\bf PC}^{(n)}} = \mathrm{ACF}_{{\bf PC}^{(n)}} \star \mathrm{ACF}_{B}$, where $ \mathrm{ACF}_{B}$ is the ACF of the beam. The FCRAO telescope beam is well approximated by a Gaussian beam of width $\sigma_B$:

\begin{equation}
B({\bf r}) = \frac{1}{2\pi \sigma_B^2} \: \exp\left(-\frac{\left|{\bf r} \right|^2}{2\sigma_B^2}\right)
\end{equation}

\noindent such that the corresponding normalized ACF is

\begin{equation}\label{eq1}
\mathrm{ACF}_{B}(\ell)  = \exp \left(-\frac{\ell^2}{4 \sigma_B^2} \right)
\end{equation}

\begin{figure*}[ht]
\centering
\includegraphics[width = 15cm]{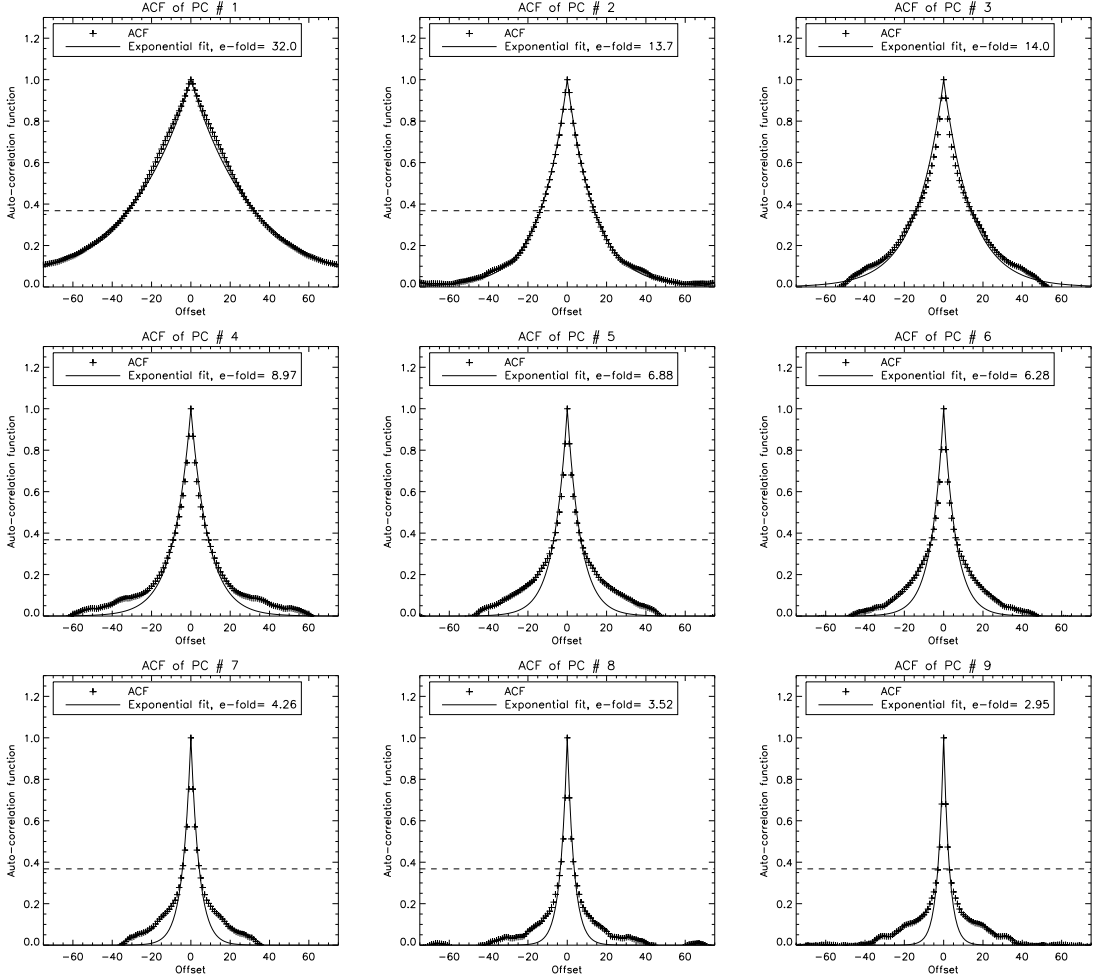}
\caption{One-dimensional ACFs of the first nine principal components (1$^{st}$ to 9$^{th}$) of one snapshot of the hydrodynamic simulation with solenoidal forcing (crosses). The solid line indicates the best exponential fit to the ACF. The $e$-fold length (in pixels) is indicated in each panel. An exponential ACF describes the ACF of the principal components of a turbulent field very accurately for orders $\geq$ 1. }
\label{exp_acf}
\end{figure*}

\indent The ACF of a turbulent field and its principal components can be well approximated by an exponential ACF \citep{brunt99, brunt02a}: $\mathrm{ACF}_{{\bf PC}^{(n)}}(\ell) = e^{-\frac{\ell}{\lambda^{(n)}}}$, where $\lambda^{(n)}$ is the true spatial scale corresponding the $n^{th}$ principal component. This is shown in Figure~\ref{exp_acf}, where the crosses represent the (true) ACF of the principal components of one time snapshot of the hydrodynamic simulation (with solenoidal forcing). The solid line indicate the exponential function with the same e-fold length as calculated in the ACF (indicated in units of pixels in the legend). An exponential function fits the ACF well, and seems to be a reasonable assumption. Note that only the core of the ACF (above the $1/e$ level) matters for the calculation of the e-fold length, and thus the fact that an exponential function does not fit the wings of the ACF is unimportant. Combining the exponential form of the true ACF of the principal components and the relation between the ACF of the observed principal components and the true principal components, the un-normalized ACF of the observed principal components is:

\begin{equation}
\mathrm{ACF}^{un}_{{\widetilde{\bf PC}}^{(n)}}(\ell) = \int_{x=0}^{\infty}{\exp\left(-{\frac{x}{\lambda^{(n)}}}\right)\exp\left(-\frac{(\ell-x)^2}{4\sigma_B^2}\right)dx}  +  \int_{x=0}^{\infty}{\exp\left(-{\frac{x}{\lambda^{(n)}}}\right)\exp\left(-\frac{(\ell+x)^2}{4\sigma_B^2}\right)dx}  
\end{equation}

\noindent The two integrands, $I_-$ and $I_+$, can be factored as:

\begin{equation}
I_{\pm} =  \mathrm{exp} \left [ -\frac{1}{4\sigma_B^2} \left (x+ 2 (\frac{\sigma_B^2}{\lambda^{(n)}} \pm \frac{\ell}{2}) \right )^2 \right ]  \mathrm{exp} \left [  \frac{1}{\sigma_B^2} \left ( \frac{\sigma_B^2}{\lambda^{(n)}} -\frac{\ell}{2} \right )^2 \right ]   \mathrm{exp} \left ( -\frac{\ell^2}{4\sigma_B^2} \right )
\end{equation}

\noindent We then make the change of variable $y = \left ( x+ 2(\sigma_B^2/\lambda^{(n)}) \pm \ell/2) \right ) / (2\sigma_B)$, and we recognize the error function (erf):


\begin{equation}
\int_a^{\infty}{e^{-y^2} dy} = \frac{\sqrt{\pi}}{2}\left( 1-\mathrm{erf}(a)\right )
\end{equation} 

\noindent so we get:

\begin{eqnarray}
\mathrm{ACF}^{un}_{{\widetilde{\bf PC}}^{(n)}}(\ell) = \sqrt{\pi}\sigma_B  \Biggl \{ \left ( 1-\mathrm{erf}\left( \frac{\sigma_B}{\lambda^{(n)}} - \frac{\ell}{2\sigma_B} \right ) \right )   \mathrm{exp} \left [  \left ( \frac{\sigma_B}{\lambda^{(n)}} -\frac{\ell}{2\sigma_B} \right )^2 \right ]   \mathrm{exp} \left ( -\frac{\ell^2}{4\sigma_B^2} \right )  \Biggr. \nonumber \\
+ \left ( 1-\mathrm{erf}\left( \frac{\sigma_B}{\lambda^{(n)}} + \frac{\ell}{2\sigma_B} \right ) \right )  \mathrm{exp} \left [  \left ( \frac{\sigma_B}{\lambda^{(n)}} +\frac{\ell}{2\sigma_B} \right )^2 \right ]   \mathrm{exp} \left ( -\frac{\ell^2}{4\sigma_B^2} \right ) \Biggr \}
\end{eqnarray}

\noindent Then we estimate $\mathrm{ACF}^{un}_{{\widetilde{\bf PC}}^{(n)}}(0)$:

\begin{equation}
\mathrm{ACF}^{un}_{{\widetilde{\bf PC}}^{(n)}}(0) = 2\sqrt{\pi}\sigma_B \Bigg(1- \mathrm{erf} \left ( \frac{\sigma_B}{\lambda^{(n)}} \right ) \Bigg) \mathrm{exp} \left [ (\frac{\sigma_B}{\lambda^{(n)}})^2 \right ]
\end{equation}

\noindent and finally get the ACF of the observed principal components normalized so that $\mathrm{ACF}_{{\widetilde{\bf PC}}^{(n)}}(0) = 1$:

\begin{eqnarray}
\mathrm{ACF}_{{\widetilde{\bf PC}}^{(n)}}(\ell) = \frac{1}{2\left(1-\mathrm{erf}(\frac{\sigma_B}{\lambda^{(n)}}) \right)} \Biggl \{   \Biggr. \nonumber \\
\left(1-\mathrm{erf}\left( \frac{\sigma_B}{\lambda^{(n)}} - \frac{\ell}{2\sigma_B}\right)\right) \mathrm{exp}\left( - \frac{\ell}{\lambda^{(n)}} \right) +  \nonumber \\
\Biggl.   \left(1-\mathrm{erf}\left( \frac{\sigma_B}{\lambda^{(n)}} + \frac{\ell}{2\sigma_B} \right) \right) \mathrm{exp}\left( \frac{\ell}{\lambda^{(n)}} \right) \Biggr \}
 \label{scale_corr}
\end{eqnarray}

\begin{figure}[ht]
\centering
\includegraphics[width = 9cm]{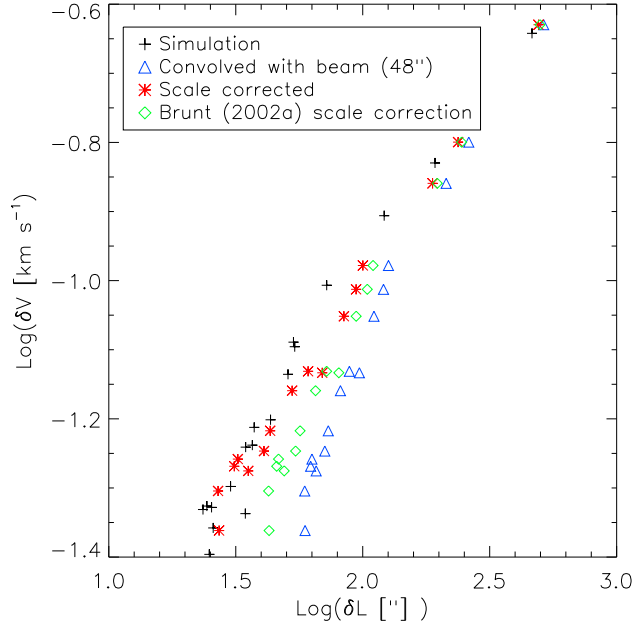}
\caption{PCA pseudo structure function derived from a snapshot PPV cube of hydrodynamic simulations (black crosses), from the same simulated cube convolved with a Gaussian beam of FWHM 48" (blue triangles), and from the beam-convolved simulated cube after scale correction (red stars). For comparison, the scales corrected with the prescription from \citet{brunt02a} are shown as green diamonds. The spatial and velocity scales were rescaled to physical units assuming a 22" grid and a temperature of 10 K (sound speed of 0.2 \kmsn). }
\label{plot_scale_corr}
\end{figure}

\noindent In order to relate the observed and true spatial scales, we need to determine for what value of $\ell$ the auto-correlation function of the principal components, $\mathrm{ACF}_{{\widetilde{\bf PC}}^{(n)}}(\ell)$, falls by one $e$-fold as a function of the true spatial scale $\lambda^{(n)}$. For true spatial scales $\lambda^{(n)}$ ranging from 5" to 5000" (sampled every 5"), we computed $\mathrm{ACF}_{{\widetilde{\bf PC}}^{(n)}}(\ell)$ according to Equation~\ref{scale_corr} and determined the corresponding ``observed'' $e$-fold length $\lambda^{(n)}_{\mathrm{obs}}$. We then constructed a look-up table relating $\lambda^{(n)}$ and $\lambda^{(n)}_{\mathrm{obs}}$, which is the desired scale correction. For every spatial scale detected in the GRS molecular clouds, the scale correction was applied by finding the two closest $\lambda^{(n)}_{\mathrm{obs}}$  values in the look-up table, and interpolating the corresponding true scale $\lambda^{(n)}$ accordingly. \\
\indent We used the hydrodynamic simulations presented in Section~\ref{hydro_sim_section} to test our scale correction. We convolved one of the simulated \CO data cubes generated from hydrodynamic simulations with a Gaussian beam of FWHM width 48", and applied PCA to the resulting beam-smeared simulated PPV cube, with and without scale correction. The result is shown in Figure~\ref{plot_scale_corr}. The black crosses represent the scales detected in the simulations without convolution with the beam (i.e., the true scales). The blue triangles represent the scales detected in the simulations after convolution with the beam, but without any scale correction. Finally, the red stars indicate the scales detected in the simulations convolved with the beam after scale correction. For comparison, the green diamonds show the scales corrected with the (approximate) prescription from \citet{brunt02a}. The power-law nature and the exponent of the PCA pseudo structure function is only recovered after correcting the spatial scales for the convolution with the telescope beam. The exact analytical scale correction presented here provides more accurate results than the prescription from \citet{brunt02a}, which is based on an approximation. Last, we emphasize that, when applying this scale correction to real observations of molecular clouds, only spatial scales above the resolution limit (as detected {\it before} the scale correction) must be taken into account. Spatial scales below the resolution limit correspond to noise and must be excluded from the analysis. In the following, we choose to use the FWHM of the beam as the resolution limit. This is justified in Section \ref{resolution_section}. The changes in PCA slopes caused by different choices of resolution limit (e.g., 3$\sigma$, $2 \times \mathrm{FWHM}$) are also investigated in Section \ref{resolution_section}.

\subsection{Uncertainties in the PCA results}
\indent The uncertainty in the spatial scales detected by PCA stems from the uncertainty in the distance. The error of the kinematic distances of the GRS clouds were estimated in \citet{RD2009} and are propagated here. The finite size of the pixels also contributes to the uncertainty on the spatial scales. Specifically, the error on the spatial scale is given by $\sigma_{\delta_\ell}^2 = \sigma_d^2 \times \delta \theta^2 + \left ( \theta_{pix}/2 \times d \right)^2$, where $\sigma_d$ is the error on the distance, $\theta_{pix}$ is the angular size of a pixel, and $\delta \theta$ is the angular scale detected by PCA. The uncertainty on the velocity scales was set to half the velocity resolution (0.1 \kms for the GRS).

\subsection{Results}
\begin{figure*}[ht]
\centering
\includegraphics[width = 15cm]{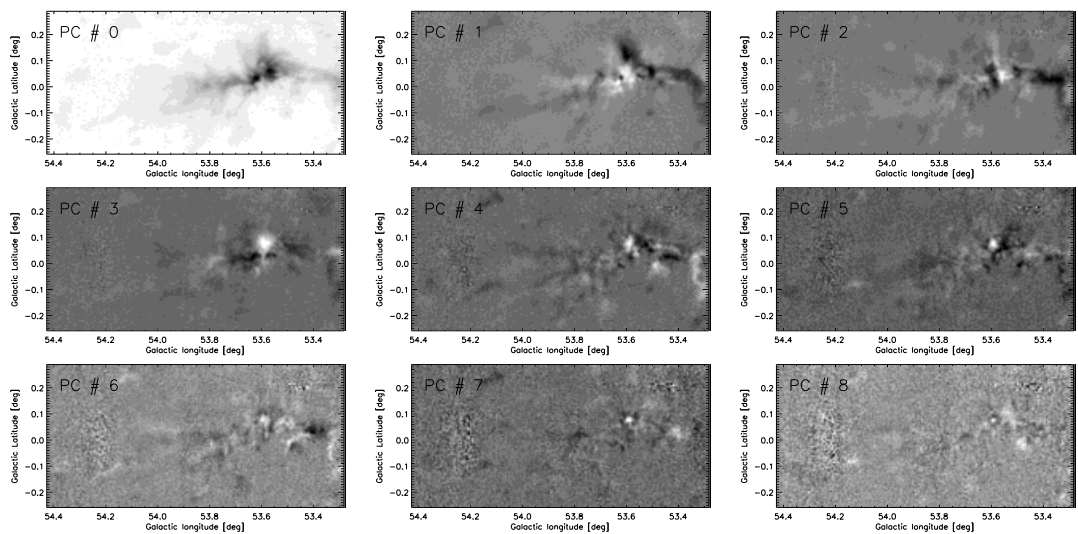}
\caption{Nine first principal components for molecular cloud GRSMC G053.59+00.04, randomly selected from our sample of 367 molecular clouds from the Galactic Ring Survey}
\label{pc_cloud}
\end{figure*}

\begin{figure}[ht]
\centering
\includegraphics[width = 9cm]{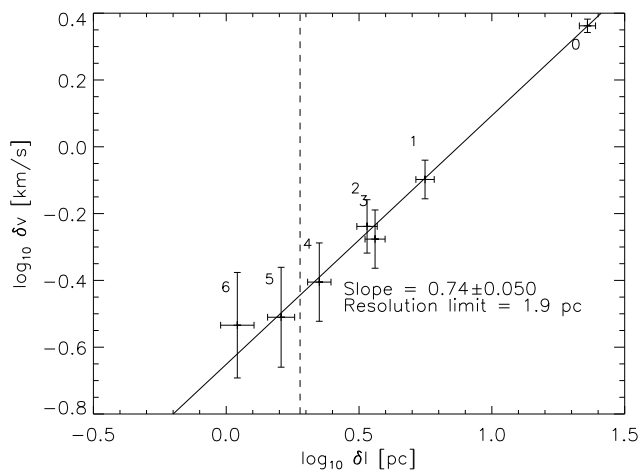}
\caption{PCA pseudo structure function for molecular cloud GRSMC G053.59+00.04. The order of the principal component for each pair of spatial and spectral scales is indicated next to each data point. The vertical dashed line shows the resolution limit. Scales detected in the 5$^{\mathrm{th}}$ and 6$^{\mathrm{th}}$ are smaller than the resolution limit after scale correction, but above it before the correction and thus need to be included in the fit. The solid line represents a power-law fit, the slope of which is indicated in the Figure. }
\label{scales_cloud}
\end{figure}

\begin{figure*}[ht]
\centering
\includegraphics[width = 15cm]{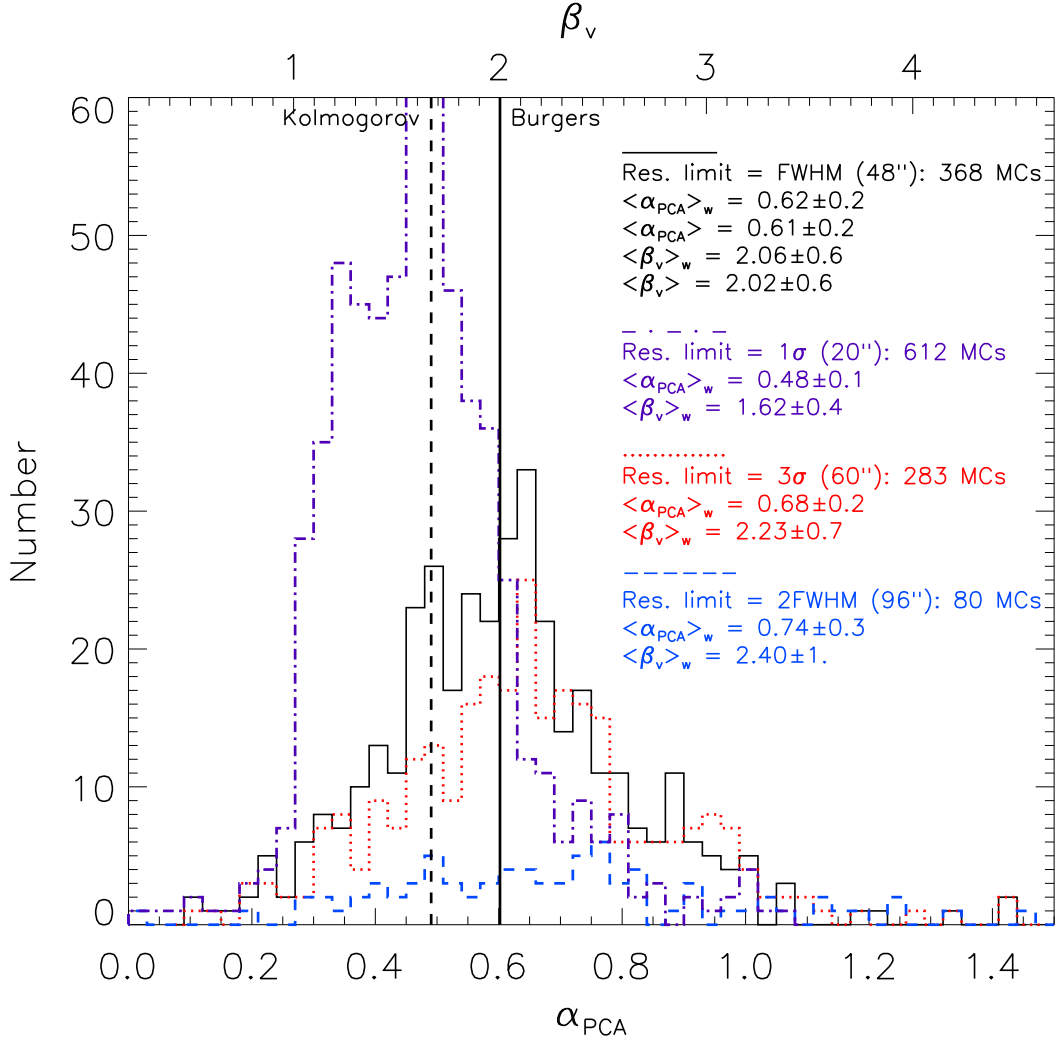}
\caption{Histograms of the slope of the PCA pseudo structure function obtained from GRS clouds, and the exponent $\beta_v$ of the turbulent spectrum obtained from the calibration derived from fBms with purely lognormal PDFs.  The errors in the legend correspond to the standard deviation of the distributions. The black histogram was derived using the FWHM of the beam as the resolution limit (fiducial case). The purple, red, and blue histograms show the histogram of $\alpha_{PCA}$ derived with resolution limits defined as the 1$\sigma$, 3$\sigma$ and 2$\times$ FWHM widths of the beam respectively. The corresponding mean PCA slopes and $\beta_v$ are also indicated for each case.}
\label{pca_histos}
\end{figure*}

\begin{figure}[ht]
\centering
\includegraphics[width = 9cm]{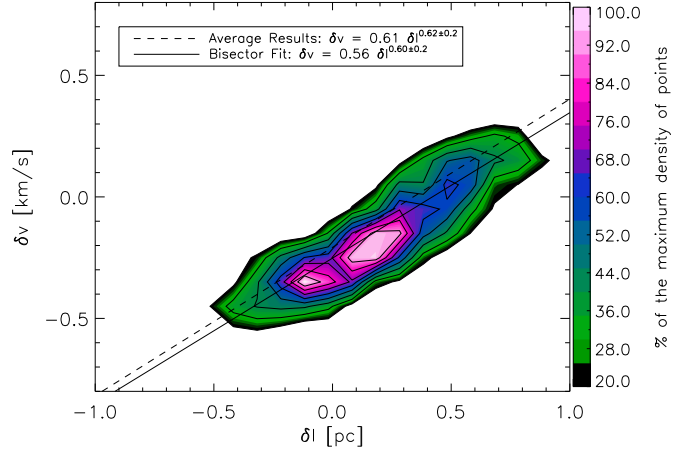}
\caption{Composite PCA pseudo structure function (composed of all the spatial and velocity scales detected in all 367 GRS molecular clouds) shown as a density of points. The dashed line indicates a power law of slope 0.62, the average slope of the PCA pseudo structure function in the GRS sample, while the solid line shows a bisector fit with slope $\alpha_{PCA}$ $=$ 0.6.}
\label{plot_cloud_all_scales}
\end{figure}

\indent PCA was applied to molecular clouds identified in the GRS for which distances are available from \citet{RD2009}. Out of the 750 molecular clouds for which distances were available, 383 did not exhibit a large-enough spatial dynamic range to allow the detection of five or more spatial and velocity scales. Consequently, a robust power law could not be fitted to the resulting PCA pseudo structure function for this sample of clouds. Our sample of GRS clouds thus contains 367 objects. Figures~\ref{pc_cloud} and~\ref{scales_cloud} show an example of PCA results for a particular cloud, GRSMC G053.59+00.04. In Figure~\ref{pc_cloud}, the 0$^{th}$ principal component simply shows the integrated intensity of the cloud and thus provides information on its overall structure. In Figure~\ref{scales_cloud}, only spatial scales above the resolution limit (before scale correction) are shown. A power law was fitted to the PCA pseudo structure function, yielding $\alpha_{PCA}$ $=$ 0.74$\pm$0.05.\\
\indent The black histogram in Figure~\ref{pca_histos} shows the histogram of the slope of the PCA pseudo structure function obtained from the GRS clouds. The mean PCA slope is $\langle\alpha_\mathrm{PCA}\rangle=0.61\pm0.2$, where the error bar reflects the standard deviation of the distribution. To reduce the effects of outliers, we also computed the average of $\alpha_{PCA}$ weighted by the inverse of the reduced $\chi^2$ of the power-law fit to the PCA structure function, and obtained a weighted average $\langle \alpha_{PCA} \rangle_w$ $=$ 0.62$\pm$0.2. This value of $\alpha_{PCA}$ is in good agreement with the PCA slope obtained for molecular clouds located in the Outer Galaxy \citep[][$\langle \alpha_{PCA} \rangle$ $=$ 0.62$\pm$0.11]{brunt02b}. A power law of slope 0.62 also fits well the composite structure function, composed of all the spatial and velocity scales detected in all the clouds (see Figure~\ref{plot_cloud_all_scales}). A bisector fit to the PCA composite structure function shown in Figure~\ref{plot_cloud_all_scales} yields $\alpha_{PCA}$ $=$0.60$\pm$0.2. \\

\subsection{Estimation of the density dispersion of GRS molecular clouds}\label{sigma_s_estimation}
\begin{figure}[ht]
\centering
\includegraphics[width = 9cm]{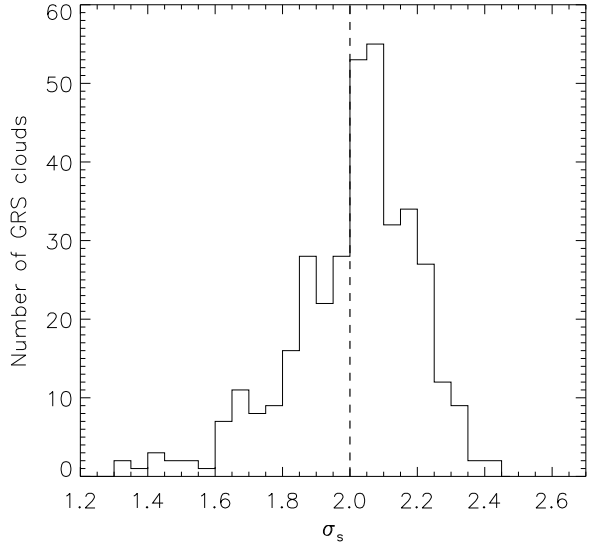}
\caption{Histogram of the log-density dispersion of GRS molecular clouds estimated from Equation \ref{sigma_s_equation}, based on their 1D velocity dispersion derived in \citet{RD2010}. }
\label{histo_grs_sigma_s}
\end{figure}

\indent Since the measured value of $\alpha_{PCA}$ is unstable above
$\sigma_{s} \approx 2$, it is worthwhile trying to estimate plausible
values of $\sigma_{s}$ that may be present in the GRS cloud
sample, to gauge the possible effects of high density dispersion on our results.
Models of driven turbulence suggest that the density
dispersion is related to the 3-dimensional rms Mach number ($\mathcal{M}$) as
follows: $\sigma_{n}/\langle~n~\rangle=b \mathcal{M}$, where $b$ is a
constant depending on the nature of the turbulent driving. For
solenoidal forcing, $b \approx 1/3$ \citep{price11}, while for compressive forcing, $b \approx 1$ \citep{fed08}.
There are very few observationally-determined values of $b$,
but existing measurements favor $b \approx 0.5$ \citep{brunt10},
indicating a mixture of solenoidal and compressive forcing.\\
\indent Assuming a lognormal PDF,  so that $\sigma_s = \sqrt{\ln(1+(\sigma_{n}/\langle n \rangle)^{2}}$, and with a specified kinetic temperature, $T$ 
and mean molecular mass, $m$, we can derive a relation between the 1-dimensional velocity standard deviation, $\sigma_{v,1D}$,
and $\sigma_{s}$ as follows:
\begin{equation}\label{sigma_s_equation}
\sigma_{s} = \sqrt{ \ln \left [ 1 + 24.69 \left[\frac{\sigma_{v,1D}}{1 \mathrm{km s}^{-1}}\right]^{2} \left[\frac{b}{0.5}\right]^{2} \left[\frac{T}{10K} \right]^{-1} \right ]} 
\end{equation}
\noindent Here, we have used $\mathcal{M}=\sigma_{v,3D}/c_{s}$, where $c_{s} = \sqrt{kT/m}$ is
the sound speed, and have assumed a mean molecular mass of 2.72 times
the mass of a hydgrogen atom \citep{hildebrand83} and taken $b = 0.5$ \citep{brunt10} and $T = 10~$K \citep{RD2010} as reference points. The choice of a kinetic temperature of 10 K is motivated by Fig.~6 in \citet{RD2010}, where the maximum excitation temperature in a molecular cloud occurs in the densest regions that are closest to LTE, and should reflect the actual kinetic temperature of the gas. We have also assumed isotropy, so that the 3-dimensional velocity standard deviation is $\sigma_{v,3D}~=~\sqrt{3}\sigma_{v,1D}$.\\
\indent Values of $\sigma_{v,1D}$ for the GRS cloud sample have already
been measured by \citet{RD2010}. We have converted these
measurements into estimates of $\sigma_{s}$ for the sample of 367
clouds analyzed here, and the histogram of the resulting $\sigma_{s}$
values is shown in Figure~\ref{histo_grs_sigma_s}. The histogram peaks near
$\sigma_{s} = 2.1$, with a tail extending to $\sigma_{s} \approx 2.4$.
Comparison of the $\sigma_{s}$ histogram with the HD results in Figure~\ref{plot_alpha_sigma_n} 
suggests that a minor overestimation of $\alpha_{PCA}$ may be present
in some clouds due to extreme density fluctuations (up to $\sim +0.1$). 
In general though, as long as $b = 0.5$ and $T = 10~$K
reasonably represent the conditions in the GRS clouds, then
we conclude that extreme density fluctuations have a relatively
minor impact on our measured $\alpha_{PCA}$. While it is unlikely
that kinetic temperatures are below 10~K, if extreme compressive forcing
is common then the $\sigma_{s}$ values will be a little higher
than represented in Figure~\ref{histo_grs_sigma_s} (but note that $\sigma_{s}$
varies only slowly with $b$ due to the square-root of a logarithm
dependence and this may be countered by raised kinetic temperatures).


\subsection{Turbulent spectrum of GRS molecular clouds from PCA}

\indent Due to the large dispersion of $\alpha_{PCA}$, PCA  provides a coarse measurement of  $\beta_v$ for any individual cloud. However, when considering the ensemble average, it is  a reliable statistical measure of the exponent of the turbulent spectrum. Applying the PCA calibration derived in Section \ref{calibration_lognormal} based on non-intermittent fBms with $\sigma_s$ $\leq$ 2, the mean value of the PCA slope (0.62$\pm$0.2) corresponds to $\langle \beta_v \rangle=2.06\pm0.6$, where the error bar reflects the standard deviation (see Figure~\ref{pca_histos}). The large standard deviation reflects not only the uncertainty on the derivation of $\alpha_{PCA}$, but potentially also intrinsic variations of the turbulent spectrum between different molecular clouds, due to varying star forming activities, different sources of forcing  (e.g., solenoidal versus compressive) and driving scales, and a range of Mach numbers \citep{klessen01, ossenkopf02, brunt09, fed09b}.   \\
\indent As pointed out before, intermittency and high density dispersion can introduce significant deviations compared to predictions from fBms. Since these effects are likely to play a significant role in molecular clouds, as shown by the HD simulations, we need to also compare the results of PCA applied to GRS molecular clouds with HD simulations. The average slope of the GRS PCA pseudo structure functions  (0.62$\pm$0.2) is in excellent agreement ($<$ 1$\sigma$) with the PCA slope derived from the spectral maps generated from solenoidally forced HD simulations ($\langle \alpha_{PCA} \rangle$ $=$ 0.65$\pm$0.05), and in marginal agreement (2$\sigma$) with the compressively forced simulations ($\langle \alpha_{PCA} \rangle$ $=$ 0.76$\pm$0.07), which likely exhibit a higher density dispersion than the average GRS molecular cloud, as demonstrated in Section \ref{sigma_s_estimation}. For these two cases of turbulence forcing, the exponents of the energy spectrum measured in the velocity fields are $\beta_v$ $=$ 1.86$\pm$0.05 and $\beta_v$ $=$ 1.94$\pm$0.05, respectively. Therefore, accounting for intermittency and density dispersion effects yields a spectral energy index of $\beta_v$ $\simeq$ 1.9 for the GRS molecular clouds.  This range of values correspond to log-Poisson (intermittent, compressible) turbulence \citep{she94, boldyrev02a, boldyrev02b, schmidt08}, but is also consistent with Burgers turbulence within the errors.  \\
\indent For the HD simulations used here, \citet{schmidt08} showed that the relation between the scaling exponents $\zeta(p)$ of the structure functions of orders $p$ $=$ 1--5 are consistent with a log-Poisson model, for which $\zeta(p)/\zeta(3)$ $=$ $(1-\Delta)\frac{p}{3} + C(1-(1-\frac{\Delta}{C})^{\frac{p}{3}})$, where $\Delta$ and $C$ are the scaling exponent (or second order structure function exponent) and co-dimension of the dominant dissipative structures, respectively. Both $C$ and $\Delta$ depend on the degree of intermittency of the flow. For the HD simulations discussed here, and assuming $\Delta$ $=$ 1 (the dominant dissipative structures are shocks that obey Burgers turbulence scaling relations), \citet{schmidt08} found $C$ $=$ 1.1 and 1.5 for the compressive and solenoidal forcing cases respectively (i.e., the dominant dissipative structures are 2D shocks).  In contrast, Burgers turbulence predicts an exponent $\zeta(p)$ $=$ 1 for  $p\geq$1, inconsistent with the scaling exponents of the structure functions in the HD simulations. Hence, if the HD simulations are an accurate model of molecular clouds, the coincidence between the exponents of the PCA pseudo structure functions derived from the HD simulations and the GRS molecular cloud catalog suggests that turbulence in molecular clouds is best described by a hierarchical, intermittent log-Poisson turbulence model with 2D shocks as the most dissipative structures. However,  we cannot formally distinguish between log-Poisson models with 2D shocks as singular dissipative structures and Burgers turbulence in the GRS molecular cloud sample due to 1) the large uncertainty in the average exponent of the energy spectrum derived from the GRS cloud sample and 2) the fact that the velocity field and structure functions are not observable.


\subsection{The choice of "resolution limit"}\label{resolution_section}
\begin{figure}[ht]
\centering
\includegraphics[width = 9cm]{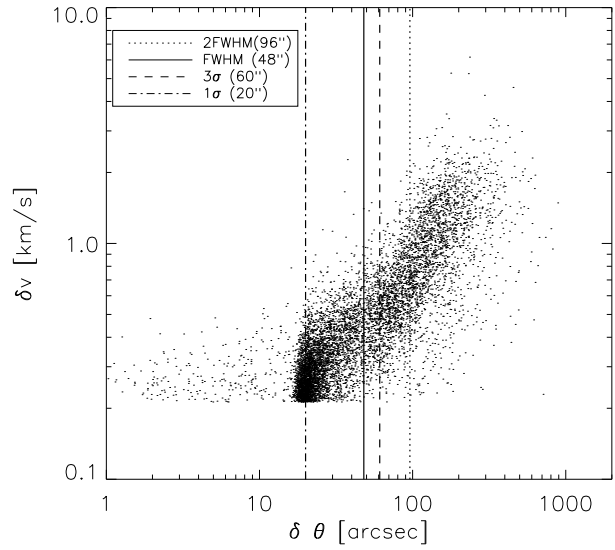}
\caption{Composite PCA pseudo structure function for GRS molecular clouds identified in the GRS, with the spatial scales left as angular scales uncorrected for beam smearing. The vertical lines indicate several characteristic values of the gaussian beam.}
\label{plot_all_scales_angular}
\end{figure}

\indent Throughout Section \ref{application_section}, we applied a cutoff to spatial scales detected by PCA. Only spatial scales above the "resolution limit" as calculated before scale correction, with the resolution limit being defined as the FWHM of the beam (48"), were considered in the PCA pseudo structure functions of GRS molecular clouds. Scales below this threshold are considered to be the result of noise and are excluded from the analysis. However, scales between the Nyquist sampling scale and twice the FWHM of the beam probably contain contributions from both astrophysical signal and noise, and our decision to use the FWHM therefore needs to be justified. \\
\indent Figure~\ref{plot_all_scales_angular} shows the composite PCA structure function, with the spatial scales left as angular scales in units of arcsecs uncorrected for beam smearing. Different characteristic values of the gaussian beam (1$\sigma$, 3$\sigma$, FWHM, 2$\times$FWHM) are indicated by vertical lines. The progressive loss of information between spatial scales corresponding to Nyquist sampling (which is equal to the 1$\sigma$ width of the beam, or 20") and the 3$\sigma$ width of the beam (60") is seen in the angular composite PCA structure function as a progressive change of slope, the slope becoming shallower as spatial scales get closer to the Nyquist sampling limit. For scales below the Nyquist sampling limit, the PCA pseudo structure function only reflects contribution from the noise, all velocity scales are equal to the spectral sampling (0.2 km s$^{-1}$), and the PCA pseudo structure function is flat (slope zero). This progressive decrease in PCA slope with decreasing spatial scales starts between 48" and 60" (the FHWM and 3$\sigma$ width of the beam). At spatial scales corresponding to 20" (the Nyquist sampling scale and the 1$\sigma$ width of the beam), the PCA angular composite structure function is dominated by the noise. Therefore, choosing the 1$\sigma$ width of the beam as the resolution limit would result in a mean PCA slope significantly skewed by the contribution of the noise. Hence, we took a conservative approach and excluded scales above the FWHM of the beam. \\
\indent Nonetheless, we have investigated the changes in PCA slope incurred by different definitions of the resolution limit. The purple, red, and blue histograms in Figure~\ref{pca_histos} show the distributions of $\alpha_{PCA}$ and $\beta_v$ obtained from resolution limits of 20" (1$\sigma$ beam width), 60" (3$\sigma$ beam width), and 96" (2$\times$FHWM) respectively. The resulting mean values of $\alpha_{PCA}$ and $\beta_v$ become steeper as the resolution limit increases (going from 1$\sigma$ of the beam to its FHWM to 3$\sigma$ to 2$\times$FWHM). This is due to the decreasing contribution of noise as the definition of the resolution limit becomes more conservative. As expected, the mean PCA slope obtained from the 1$\sigma$ definition of resolution limit is significantly lower than for the 3 other cases because it includes a large contribution from the noise. Excluding this case, the mean values of $\alpha_{PCA}$ and $\beta_v$ calculated with different definitions of the resolution limit are within the errors of each other, and all consistent well within the errors with the HD simulations. We conclude that, as long as the resolution limit is above the FWHM of the beam, the choice of resolution limit (between FWHM, 3$\sigma$, and 2$\times$ FWHM) does not change the interpretations and the conclusions presented here.


\section{Summary and conclusion}\label{conclusion}

\indent We applied Principal Component Analysis (PCA) to synthetic Position-Position-Velocity (PPV) spectral maps generated from the density and velocity fields of solenoidally and compressively forced hydrodynamic simulations of supersonic turbulence \citep{fed08, fed09a, fed09b}, and of fractional Brownian motion simulations, in order to constrain the calibration relation between the PCA pseudo structure function and the index of the velocity spectrum of turbulence, and to examine the dependency of this relation on the density spectrum, intermittency, and density dispersion.\\
\indent We demonstrated that the calibration relation, the relation between the slope of the PCA structure function $\alpha_{PCA}$ and $\beta_v$, does not depend on the exponent of the power-law density spectrum $\beta_n$. \\
\indent For a log-density dispersion $\sigma_s$ $\leq 2$,  we do not find any dependence of the PCA calibration on the dispersion of the density PDF. We derive a PCA calibration relation, $\beta_v = 0.20\pm0.05 + (2.99\pm0.09)\alpha_{PCA}$ valid for $\sigma_s$ $\leq$ 2 and $\beta_v$ $=$ 1.2--2.6. For $\sigma_s$ $>$ 2, we find a strong dependence of the calibration between $\alpha_{PCA}$ and $\beta_v$ with $\sigma_s$. Extreme density fluctuations intermittently sample the velocity field, producing an effect similar to intermittency in the velocity field itself - i.e. mimicking discontinuous velocity jumps, although the detailed mechanism is rather different. PCA is stable below a threshold of the log-density dispersion, $\sigma_s$ $\simeq$ 2, but if real molecular clouds exceed this, then an additional overestimation factor applies to $\alpha_{PCA}$. Without knowledge of the true 3D log-density dispersions in the cloud sample, the estimation of the turbulent spectrum in molecular clouds remains uncertain. \citet{brunt10a, brunt10b} developed a method to estimate the density PDF of molecular clouds based on the 2D power spectrum, the variance, and the PDF of the 2D column density, from which the 3D density PDF can be reconstructed, even in cases where the density PDF is not lognormal. However, this method requires high fidelity measures of column density such as extinction derived from 2MASS photometry of background stars and high spatial dynamic range. Therefore, it is not readily applicable to our set of data from the Galactic Ring Survey for which the spatial dynamic range for most clouds is limited. In addition, numerical simulations predict a relation between the log-density dispersion and the Mach number \citep{price11}, but this relation also depends on the relative contribution of solenoidal and compressive modes \citep{fed09b}. An initial test of the log- density dispersion - Mach number relation has been made \citep{brunt10a, brunt10b}, and this suggests that both solenoidal and compressive forcing are important and that density dispersions are likely to be high enough that their effect on PCA is not insignificant. \\
\indent We demonstrated that intermittency in the velocity field also increases the PCA slope for a given velocity spectrum. This effect is due to a combination of the operating order of PCA (PCA traces the first order structure function exponent, $\zeta(1)$), and the variation of the ratio between $\zeta(1)$ and $\beta_v$ with intermittency. Thus, if a first-order scheme is used to measure the second order exponent $\beta_v$, then some knowledge of the level of intermittency is required. By accounting for the level of intermittency, we were able to reconcile PCA measurements between non-intermittent fBms and the HD fields.  \\
\indent We applied PCA to \CO spectral maps of 367 molecular clouds identified in the Galactic Ring Survey \citep{GRS}. We found that the average slope of the PCA pseudo structure function and the slope of the composite structure function, made of all the spatial and velocity scales derived in all the GRS clouds, are consistent with $\alpha_{PCA}$ $=$ $0.62\pm0.2$. Applying the PCA calibration obtained from fBms with $\sigma_s$ $\leq$ 2, the PCA slope obtained for GRS molecular clouds corresponds to an average turbulence spectral index of $\langle\beta\rangle=2.06\pm0.6$. However, we have shown that intermittency and density dispersion need to be taken into account.  The average PCA slope obtained for the GRS clouds is in very good agreement with the PCA slope obtained from both solenoidally and compressively forced HD simulations, albeit in better agreement  (at $<$ 1 $\sigma$) with the solenoidally forced HD simulations. This agreement suggests that  turbulence in molecular clouds, as in the HD simulations, obey log-Poisson scaling relations (intermittent, compressible turbulence) with 2D shocks as the dominant dissipative structures.


\acknowledgments{
We thank Mordecai Mark Mac Low for fruitful discussions, and the referee for a detailed and balanced report. This work was supported by NSF grant AST-0507657. The molecular line data used in this paper is from the Boston University (BU)-FCRAO GRS, a joint project of Boston University and the Five College Radio Astronomy observatory funded by the National Science Foundation under grants AST 98-00334, AST 00-98562, AST 01-00793, AST 02-28993, and AST 05-07657. CF and RSK acknowledge funding by the Baden-W\"{u}rttemberg Stiftung via contract P-LS-SPII/18 in the program {\em Internationale Spitzenforschung II}, the Max-Planck-Institute, the IMPRS-A and the HGSFP at Heidelberg University (funded by the Excellence Initiative of the DFG under grant GSC~129/1). CF and RSK also acknowledge financial support from the German Bundesministerium f\"{u}r Bildung und Forschung via the ASTRONET project STAR FORMAT (grant 05A09VHA). CF also received funding from the European Research Council under the European Community's Seventh Framework Programme (FP7/2007-2013 Grant Agreement no. 247060). RSK thanks for support from the collaborative research project SFB 881 {\em The Milky Way System} funded by the DFG. RSK furthermore thanks for subsidies from the FRONTIER grant of Heidelberg University sponsored by the German Excellence Initiative. CF acknowledges computational resources from HLRB~II project (grant no.~pr32lo) at the Leibniz Rechenzentrum Garching and from the J\"{u}lich Supercomputing Center (grant no.~HHD20).}

\bibliographystyle{apj}
\bibliography{bibliography_thesis_work}

\begin{table}
\begin{center}
\caption{Turbulence statistics obtained for hydrodynamic simulations}
\def\arraystretch{1.3}
\begin{tabular}{cccc}
\hline
\hline
 Symbol & Description & Solenoidal Forcing & Compressive Forcing \\
\hline
$\beta_n$  & Exponent of the density spectrum & $\phantom{+}0.78\pm0.06$    & $\phantom{+}1.44\pm0.23$   \\
$\beta_v$  & Exponent of the velocity spectrum & $\phantom{+}1.86\pm0.05$    & $\phantom{+}1.94\pm0.05$   \\
$\sigma_n/\langle n \rangle$ & Standard Deviation of $n$ & 1.89$\pm$0.09  & 5.86$\pm$0.96 \\
$\langle s \rangle$  & Mean of $s = ln(n/\langle n \rangle)$  & $-$0.83$\pm$0.05 & $-$3.40$\pm$0.43 \\
$\sigma_s$ & Standard Deviation of $s$ & 1.32$\pm$0.06  & 3.04$\pm$0.24 \\
$\mathcal{S}_s$ & Skewness of  $s$ & -0.1$\pm$0.11 & $-$0.26$\pm$0.20\\
$\mathcal{K}_s$ & Kurtosis of $s$ & 3.03$\pm$0.17 & 2.91$\pm$0.43 \\

\hline
\end{tabular}
\label{stat_hydro_table}
\end{center}
\end{table}

\begin{table}
\begin{center}
\caption{Density dispersion of the fBm density fields with varying $\beta_n$ and gaussian density PDF}
\def\arraystretch{1.3}
\begin{tabular}{cccc}
\hline
\hline
\hline
$\beta_n$ & $\sigma_n/\langle n \rangle$ & $\sigma_s$ \\
    	        0.600000  & 0.20 & 0.21\\
     0.800000  & 0.18 & 0.19\\
      1.00000  & 0.19 & 0.21\\
      1.20000  & 0.21 & 0.22\\
      1.40000  & 0.22 & 0.23\\
      1.60000  & 0.23 & 0.25\\
      1.80000  & 0.23 & 0.26\\
      2.00000  & 0.25 & 0.28\\
      2.20000  & 0.28 & 0.32\\
      2.40000  & 0.27 & 0.33\\
      2.60000  & 0.36 & 0.39\\
      2.80000  & 0.34 & 0.40\\
      3.00000  & 0.34 & 0.40\\
      3.20000  & 0.32 & 0.41\\
      3.40000  & 0.30 & 0.41\\

\hline
\end{tabular}
\label{table_density_disp}
\end{center}
\end{table}

\begin{table}
\begin{center}
\caption{Moments of the density PDF and spectral index, $\beta_n$,  for the fBms with lognormal density PDF and varying density dispersion, $\sigma_n$}
\def\arraystretch{1.3}
\begin{tabular}{cccccc}
\hline
\hline
\hline
$\sigma_n/\langle n \rangle $ & $\langle s \rangle $ & $\sigma_s$ & $\mathcal{S}_s$ & $\mathcal{K}_s$  & $\beta_n$  \\
\hline
0.2 &   -0.02   &   0.20  &  0.03   &   3.02   &  1.04\\
2.00 &    -0.81  &   1.27  &  0.04   &   2.95   &  0.63\\
4.09 &     -1.42 &   1.68  &  0.01   &    2.98  &  0.37\\
5.92 &    -1.83  &   1.90  & 0.04 &  2.97    &  0.22\\
8.01 &     -2.11 &   2.04  & 0.02    &   2.97   &  0.16\\
10.02 & -2.33 & 2.15 &  0.04 & 2.97 & 0.22 \\
12.55 &   -2.47 &  2.23 &  -0.01    & 2.99   &   0.11   \\
14.65 & -2.62 & 2.30 & 0.00 & 2.97 &    -0.05    \\
19.33 &     -3.07  &    2.45 & 0.01   & 3.02&   -0.17\\
\end{tabular}
\label{table_beta_n}
\end{center}
\end{table}



\end{document}